\documentclass[]{article}
\usepackage{graphicx}
\usepackage{epsfig}
\usepackage{epstopdf}
\usepackage{authblk}
\usepackage{subfigure}
\usepackage{wrapfig}
\usepackage[T1]{fontenc}
\usepackage[latin9]{inputenc}
\usepackage{units}
\usepackage{amsmath}
\usepackage{amssymb}
\usepackage{esint}
\usepackage{float}
\usepackage{longtable,tabularx}
\usepackage{color}
 \title{{\normalsize under consideration for publication in AIAA Journal} \vspace {7mm} \\
 Assessment of the impact of two-dimensional wall deformations' shape on high-speed boundary layer disturbances}

\author[1]{Jeremy Sawaya }
\author[1]{Vasileios Sassanis }
\author[1]{Sofia Yassir }
\author[1]{Adrian Sescu}
\author[2]{Miguel Visbal}

\affil[1]{Department of Aerospace Engineering, Mississippi State University,  MS 39762}
\affil[2]{Air Force Research Laboratory, Wright-Patterson Air Force Base, OH 45433}

\begin{document}

\date{}

\maketitle

\begin{abstract}

Previous experimental and numerical studies showed that two-dimensional roughness elements can stabilize disturbances inside a hypersonic boundary layer, and eventually delay the transition onset. The objective of this paper is to evaluate the response of disturbances propagating inside a high-speed boundary layer to various two-dimensional surface deformations of different shapes. We perform an assessment of the impact of various 2D surface non-uniformities, such as backward or forward steps, combinations of backward and forward steps, wavy surfaces, surface dips, and surface humps. Disturbances inside a Mach 5.92 flat-plate boundary layer are excited using periodic wall blowing and suction at an upstream location. The numerical tools consist
 of a high-accurate numerical algorithm solving for the unsteady, compressible form of the Navier-Stokes equations in curvilinear coordinates. Results show that all types of surface non-uniformities are able to reduce the amplitude of boundary layer disturbances to a certain degree. The amount of disturbance energy reduction is related to the type of pressure gradients that are posed by the deformation (adverse or favorable). A possible cause (among others) of the disturbance energy reduction inside the boundary layer is presumed to be the result of a partial deviation of the kinetic energy to the external flow, along the discontinuity that is generated by the wall deformation. 

\end{abstract}



\section{Introduction}

Surface imperfections are important disturbing factors in boundary layer transition, and it is known either from experiments (Gregory et al. \cite{Gregory}, Drake et al. \cite{Drake}, Duncan et al. \cite{Duncan}) or numerical simulations (Choudhari and Fischer \cite{choudhari1}, Yoon et al. \cite{Yoon}, Muppidi and Mahesh \cite{Muppidi}, Iyer et al. \cite{Iyer}, Brehm et al. \cite{Brehm}, Duan and Choudhari \cite{Duan}, Subbareddy et al. \cite{Subbareddy}, Rizzetta and Visbal \cite{Rizzetta5}, Sescu et al. \cite{Sescu1,Sescu5}, Chaudhry et al. \cite{Chaudhry}) that they can have a significant impact on the boundary layer receptivity and transition. Direct numerical simulations showed that small steps may impact the transition onset, depending on the type and height of the step, as well as the flow conditions. 

The interest in studying the effect of surface imperfections on the transition in supersonic and hypersonic boundary layers has been revitalized in the recent years. The transition at supersonic speeds is sensitive to the shape and height of surface imperfections and the Reynolds number as in the incompressible regime. In addition, it is also dependent on the Mach number, free-stream temperature, the thermal boundary conditions at the wall, and shock waves that may develop due to the presence of the imperfections. The way the latter impacts the transition in high-speed boundary layers is still unclear. Most of the studies involving surface imperfections looked at isolated roughness elements of different shapes (Fong et al. \cite{Fong1,Fong2,Fong3,Fong4}, Duan et al. \cite{Duan2}, Park and Park \cite{Park}, Mortensen and Zhong \cite{Mortensen}, Bountin et al. \cite{Bountin}). A comprehensive review about the effect of different roughness elements on hypersonic boundary layers can be found in Schneider \cite{Schneider}. 

Acoustic waves were found to be very effective in exciting high-speed boundary layers because the phase speed of the acoustic waves synchronizes with the phase speed of the first modes that correspond to the lower branch of the neutral stability curve. There are numerous studies involving the interaction of acoustic waves with supersonic boundary layers (e.g., Mack \cite{Mack}, Gaponov \cite{Gaponov1}, Gaponov and Smorodsky \cite{Gaponov2}, Fedorov and Khokhlov \cite{Fedorov1,Fedorov2}, Sakaue et al. \cite{Sakaue}, Fedorov \cite{Fedorov3}). In some of these studies, it was found that acoustic waves are very effective in exciting disturbances inside the boundary layer with amplitudes that become much larger than those in the free-stream, but this happens only above some critical Reynolds number as in the incompressible regime. Other studies (Fedorov and Khokhlov \cite{Fedorov1,Fedorov2}, Sakaue et al. \cite{Sakaue}, Fedorov \cite{Fedorov3}) were concerned about the generation of the first and second modes in the vicinity of the leading edge. The effect of all types of waves, i.e. slow and fast acoustic waves, vorticity waves and entropy waves, on supersonic boundary layers was studied and reported in a suite of papers by Balakumar \cite{Balakumar1,Balakumar2,Balakumar3}. The generation and the evolution of three-dimensional disturbances induced by slow and fast acoustic disturbances and isolated roughness elements in a supersonic boundary layer over flat plates and wedges were numerically investigated by solving the full three-dimensional Navier-Stokes equations. It was found that instability waves are generated within one wavelength of the acoustic wave from the leading edge.

Previous experimental and numerical studies (Holloway and Sterrett \cite{Holloway}, Fujii \cite{Fujii}, Fong et al. \cite{Fong1,Fong2,Fong3,Fong4}, Duan et al. \cite{Duan2}, Park and Park \cite{Park}, Mortensen and Zhong \cite{Mortensen}) showed evidence that two-dimensional roughness elements can reduce the amplitude of disturbances inside high-speed boundary layers. Holloway and Sterrett \cite{Holloway}, for example, carried out early experiments on flat plate boundary layer disturbed by roughness elements, for free stream Mach number of 4.0 and 6.0, and observed a delay in the transition for roughness elements with height smaller than local boundary layer thickness. Fujii \cite{Fujii} conducted an experimental investigation of the effect of two-dimensional roughness elements on a hypersonic Mach $7.1$ boundary layer developing over a half-angle sharp cone. It was found that a wavy surface with the wavelength equal to twice the boundary layer thickness delayed the transition onset. Duan et al. \cite{Duan2} and Fong et al. \cite{Fong1,Fong2,Fong3,Fong4} in a series of studies investigated the effect of two-dimensional roughness on the instability of the second mode (or mode S) by direct numerical simulations (DNS). Their numerical results proved that the roughness located at the downstream of the synchronization point is able to stabilize this mode. Park and Park \cite{Park} studied the effect of a two-dimensional smooth hump on linear instability of hypersonic boundary layer by using parabolized stability equations. Their results confirmed the findings of the previous studies, i.e., the mode S is stabilized by the hump when it is located in the downstream of the synchronization point, but they also found that this mode is destabilized if the hump is located in the upstream of the synchronization point. Experimental and computational work by Bountin et al. \cite{Bountin} showed that wavy surfaces lead to a considerable reduction of the spectral peak associated with the second-mode instability in a Mach 6 boundary layer.

Previous studies focused on localized disturbances propagating as wave packets inside the boundary layer. In this work, we study the effect of various 2D surface non-uniformities on both pulsed and periodic disturbances propagating inside a high-speed boundary layer (the focus is on periodic disturbances that have not received much attention previously). Different wall non-uniformities are considered here: backward or forward small steps, combinations of backward and forward small steps, wavy surfaces with the mean above or below the wall surface, surface dips, or surface humps. The numerical tool is a high-accurate solver, discretizing the unsteady, compressible, conservative form of the Navier-Stokes equations written in body-fitted curvilinear coordinates.  Velocity and temperature profiles corresponding to a compressible boundary layer are imposed at the inflow, thus avoiding the inclusion of the leading edge shock in the computation. Since this is a 2D study, there are some limitations in terms of the types of modes being considered: for example, in the 2D framework only the second mode is predominant, while the oblique first mode is not captured by the analysis. In the results section, it is found that all types of non-uniformities are capable of reducing the amplitude of boundary layer disturbances to a certain degree. It is suggested that the oblique Mach wave that is posed by the wall deformations is responsible for deviating a small portion of the kinetic energy of the disturbance to the external flow. This may be a potential cause (among others) for disturbance energy reduction in the downstream of the roughness element. The type of pressure gradient (adverse or favorable) that is posed first by the surface non-uniformity is also a factor in the reduction of the disturbance energy, among other factors, such as the location of the synchronization point with respect to the location of the roughness element, as was found in previous studies.

In section II, the governing equations and the numerical tool are briefly introduced and described. In section III, the linear stability analysis methodology is briefly discussed. Section IV is reserved to results and discussions, where various qualitative and quantitative plots are reported and discussed. Final conclusions are included in section V.

\section{Problem formulation and numerical algorithm}

\subsection{Scalings}

In this study, the governing equations consist of the full Navier-Stokes equations written in generalized curvilinear coordinates, where the spatial coordinates in the computational space are expressed in terms of the spatial coordinates in the physical space as
$\xi = \xi \left(x,y \right),
\eta = \eta \left(x,y \right)$,
where $x$, $y$ correspond to the streamwise and wall-normal directions. This transformation allows for a seamless mapping of the solution from the computational to the physical space and vice-versa. All dimensional spatial coordinates are normalized by the boundary layer thickness at the inflow boundary, $\delta^*$, i.e.,

\begin{eqnarray}\label{NS}
(x,y) = \frac{(x^*,y^*)}{\delta^*},
\end{eqnarray}
the velocity is scaled by the freestream velocity magnitude $V_{\infty}^*$,

\begin{eqnarray}\label{NS}
(u,v) = \frac{(u^*,v^*)}{V_{\infty}^*}, 
\end{eqnarray}
the pressure by the dynamic pressure at infinity, $\rho_{\infty}^* V_{\infty}^{*2}$, and temperature by the freestream temperature, $T_{\infty}^*$. Reynolds number based on the boundary layer thickness, Mach number, and Prandtl number are defined as 

\begin{eqnarray}\label{NS}
R_{\lambda} = \frac{\rho_{\infty}^* V_{\infty}^* \delta^*}{\mu_{\infty}^*}, \hspace{5mm}
Ma = \frac{V_{\infty}^*}{a_{\infty}^*}, \hspace{5mm}
Pr = \frac{\mu_{\infty}^* C_p}{k_{\infty}^*}. 
\end{eqnarray}
where $\mu_{\infty}^*$, $a_{\infty}^*$ and $k_{\infty}^*$ are freestream dynamic viscosity, speed of sound and thermal conductivity, respectively, and $C_p$ is the specific heat at constant pressure. All simulations are performed for air as an ideal gas (no species equations are considered since it is expected that the effect of chemical reactions is negligible in the development of disturbances).

\subsection{Governing equations}

We consider a hypersonic flat-plate boundary layer with a very small two-dimensional surface non-uniformity located at a certain distance from the leading edge. In conservative form, the Navier-Stokes equations are written as

\begin{eqnarray}\label{NS}
\mathbf{Q}_t
+ \mathbf{F} _{\xi}
+ \mathbf{G}_{\eta}
= \mathbf{S}.
\end{eqnarray}
where the vector of conservative variables is given by

\begin{equation}
\mathbf{Q} = \frac{1}{J} \{ 
\begin{array}{rrrrrr}
\rho,   \hspace{4mm}
\rho u_i,   \hspace{4mm}
E
\end{array}
\}^{T}, i = 1,2
\end{equation}
$\rho$ is the density of the fluid, $u_i = (u, v)$ is the velocity vector in physical space, and $E$ is the total energy. The flux vectors, $\mathbf{F}$ and $\mathbf{G}$, are given by

\begin{eqnarray}
\mathbf{F} = \frac{1}{J} \left\{ 
\begin{array}{c}
\rho U,   \hspace{4mm}
\rho u_iU + \xi_{x_i} (p + \tau_{i1}),   \hspace{4mm}
E U + p \tilde{U} +  \xi_{x_i} \Theta_i
\end{array}
\right\}^{T}, \\
\mathbf{G} = \frac{1}{J} \left\{ 
\begin{array}{c}
\rho V,   \hspace{4mm}
\rho u_iV + \eta_{x_i} (p + \tau_{i2}),    \hspace{4mm}
E V + p \tilde{V}+  \eta_{x_i} \Theta_i
\end{array}
\right\}^{T}
\end{eqnarray}
where the contravariant velocity components are given by

\begin{eqnarray}
U = \xi_{x_i} u_i ,   \hspace{4mm}
V = \eta_{x_i} u_i
\end{eqnarray}
with the Einstein summation convention applied over $i,j$, the shear stress tensor and the heat flux are given as

\begin{equation}
\tau_{ij} = \frac{\mu}{Re} \left[
\left(
\frac{\partial \xi_k}{\partial x_j}  \frac{\partial u_i}{\partial \xi_k}  +
\frac{\partial \xi_k}{\partial x_i}  \frac{\partial u_j}{\partial \xi_k}
\right)
- \frac{2}{3} \delta_{ij} \frac{\partial \xi_l}{\partial x_k}  \frac{\partial u_k}{\partial \xi_l}
\right]
\end{equation}

\begin{equation}
\Theta_{i} = 
 u_j \tau_{ij} + \frac{\mu}{(\gamma-1)M_{\infty}^2 Re Pr}
\frac{\partial \xi_l}{\partial x_i}  \frac{\partial T}{\partial \xi_l}
\end{equation}
respectively, and $\mathbf{S}$ is the source vector term. The pressure $p$, the temperature $T$  and the density of the fluid are combined in the equation of state, $p = \rho R T$ where non-chemically-reacting flows are considered ($R$ is the gas constant). The Jacobian of the curvilinear transformation from the physical space to computational space is denoted by $J$. The derivatives $\xi_x$, $\xi_y$, $\eta_x$, $\eta_y$ represent grid metrics. The dynamic viscosity $\mu$ and thermal conductivity $k$ is linked to the temperature using the Sutherland's equations in dimensionless form,

\begin{eqnarray}
\mu = T^{3/2} \frac{1 + C_1/T_{\infty}}{T+C_1/T_{\infty}}; \hspace{4mm}
k = T^{3/2} \frac{1 + C_2/T_{\infty}}{T+C_2/T_{\infty}},
\end{eqnarray}
where for air at sea level, $C_1 = 110.4 K$, $C_2 = 194 K$, and $T_{\infty}$ is a reference temperature.

A high-order numerical algorithm is employed to solve the Navier-Stokes equations, wherein the time integration is performed using a third order TVD Runge-Kutta method~\cite{liu} written in the form

\begin{eqnarray}\label{28}
\mathbf{Q}^{(0)} &=& \mathbf{Q}^n                 \nonumber \\
\mathbf{Q}^{(1)} &=& \mathbf{Q}^{(0)}+\Delta t L(u^{(0)}) \nonumber \\
\mathbf{Q}^{(2)} &=& \frac{3}{4}\mathbf{Q}^{(0)}+\frac{1}{4}\mathbf{Q}^{(1)}+\frac{1}{4}\Delta t L(\mathbf{Q}^{(1)}) \\
\mathbf{Q}^{n+1} &=& \frac{1}{3}\mathbf{Q}^{(0)}+\frac{2}{3}\mathbf{Q}^{(1)}+\frac{2}{3}\Delta t L(\mathbf{Q}^{(2)}), \nonumber
\end{eqnarray}
where $L(\mathbf{Q})$ is the residual and $\Delta t$ is the time step. The spatial derivatives are discretized using either a dispersion relation preserving scheme (Tam and Webb \cite{Tam}) or a high-resolution 9-point dispersion-relation-preserving optimized scheme of Bogey et al. \cite{Bogey3}. 
The spatial discretization scheme can be written as
$\left( \partial_x f \right)_l \simeq1/\Delta x \sum_{ j = -N }^{ M } a_j f_{l+j}$
where the coefficient are given in table \ref{ttt1}.

\begin{table}
 \begin{center}
  \caption{Weights of the centered stencils}
  \label{ttt1}
  \begin{tabular}{rrrrrrrr} \hline
       Stencil & $a_1=-a_{-1}$ & $a_2=-a_{-2}$ & $a_3=-a_{-3}$ & $a_4=-a_{-4}$ &  \\\hline
       $DRP$ &  0.77088238 &  -0.16670590 & 0.02084314 & 0 &\\
       $FDo9p$ &  0.84157012 &  -0.24467863 & 0.05946358 & -0.00765090 &  \\ \hline
  \end{tabular}
 \end{center}
\end{table}

To damp out the unwanted high-wavenumber waves from the solution, high-order spatial filters, as developed by Kennedy and Carpenter \cite{Kennedy}, are applied to all variables. Nonreflecting boundary conditions (Kim and Lee \cite{Kim}) are used at the inflow boundary and an extrapolation condition is imposed at the outflow boundary. The mean inflow conditions, consisting of velocity, density and temperature profiles, are obtained separately from a precursor two-dimensional simulation, where a Blasius type boundary condition is imposed in the upstream. A 'slice' of data from the two-dimensional flow domain is imposed at the inflow boundary of the main domain.

No slip boundary conditions for velocity and isothermal condition for temperature are imposed at the solid surface.
 Sponge layers are imposed in the proximity of the far-field boundaries, and combined with grid stretching to damp out the unwanted spurious waves; these sponge layers are set outside the flow domain since they generate unphysical solutions (Sescu et al. \cite{Sescu1}). Shock capturing techniques are needed to avoid unwanted oscillations that may propagate from potential discontinuities. In this study, we apply a shock capturing
methodology that was proven to work efficiently for high-order, nonlinear computations (Bogey et al. \cite{Bogey2}). Since in the present work high-order, central-difference schemes
are used to achieve increased resolution of the propagating disturbances, a straightforward
approach is a model which introduces sufficient numerical
viscosity in the area of the discontinuities, and negligible artificial viscosity in the
rest of the domain. A shock-capturing technique, suitable for simulations involving
central differences in space is applied, based on the
general explicit filtering framework. The technique introduces selective filtering at each grid vertex to minimize numerical oscillations, and shock-capturing in the areas where discontinuities are present (more details can be found in Bogey et al. \cite{Bogey2}).
 




\section{Linear Stability Equations}

In two-dimensional Cartesian coordinates, $x$ is defined as the streamwise direction and $y$ is defined as the wall-normal direction.  All of the velocity components are scaled by the reference velocity $V_{\infty}$, the spatial coordinates by the boundary layer thickness $\delta$, density by $\rho_{\infty}$, pressure by $\rho_{\infty} V_{\infty}^2$, time by $\delta/V_{\infty}$, and other variables are scaled by the corresponding boundary layer edge values (Malik et al. \cite{Malik}). The instantaneous velocity ($u$, $v$), pressure ($p$), temperature ($\tau$), density ($\rho$), dynamic viscosity ($\mu$), and thermal conductivity coefficient ($k$) are represented as a summation of the mean and the disturbance as

\begin{eqnarray}
u = \overline{U} + \tilde{u},  \hspace{3mm}
v = \overline{V} + \tilde{v},  \hspace{3mm}
p = \overline{P} + \tilde{p},  \hspace{3mm}
\tau = \overline{T} + \tilde{T},  \hspace{3mm}
\rho = \overline{\rho} + \tilde{\rho},  \hspace{3mm}
\mu = \overline{\mu} + \tilde{\mu},  \hspace{3mm}
k = \overline{k} + \tilde{k}.
\end{eqnarray}

The "locally parallel flow" assumption is employed here for the compressible boundary layer flow.  With this assumption in place, the mean quantities are a function of the wall-normal coordinate only, i.e.

\begin{eqnarray}
U = U(y),  \hspace{4mm}
V = V(y),  \hspace{4mm}
T = T(y),  \hspace{4mm}
\rho = \rho(y).
\end{eqnarray}
where $P$ is assumed constant across the boundary layer and equal to $1/{\gamma}M^2$, and $\rho = 1/T$.  Thus, the equation for the density disturbance $\tilde{\rho}$ becomes,

\begin{equation}
\tilde{\rho} = {\gamma}M^2\frac{\tilde{p}}{T}-\frac{\tilde{T}}{T^2}.
\end{equation}

From Sutherland equations, $\tilde{\mu}$ and $\tilde{k}$ can be expressed as

\begin{eqnarray}
\begin{array}{c}
\tilde{\mu} = \frac{d\mu}{dT}\tilde{T}, \hspace{4mm}
\tilde{k} = \frac{dk}{dT}\tilde{T}.
\end{array}
\end{eqnarray}

To derive the stability equations, the fluctuations in velocity, pressure and temperature are assumed to resemble a harmonic wave defined as 

\begin{equation}\label{a1}
[\tilde{u}, \tilde{v}] = [\hat{u}(y), \hat{v}(y)]e^{i({\alpha}x - {\omega}t)}
\end{equation}  
\begin{equation}\label{a2}
\tilde{p} = \hat{p}(y)e^{i({\alpha}x - {\omega}t)}
\end{equation}  
\begin{equation}\label{a3}
\tilde{T} = \hat{T}(y)e^{i({\alpha}x - {\omega}t)},
\end{equation}
where $\alpha$ is the wavenumber and $\omega$ is the frequency. Within the spatial stability analysis, the frequency $\omega$ is considered real, while the wavenumber $\alpha$ is a complex number to be determined; in the temporal stability analysis, the wavenumber $\alpha$ is considered real, and the frequency $\omega$ is a complex unknown. The Navier-Stokes equations are first linearized around the mean flow, resulting in a set of equations for disturbances. Then, the ansatz (\ref{a1})-(\ref{a3}) is plugged into the disturbance equations to obtain the following system of ordinary differential equations (which forms an eigenvalue problem)

\begin{equation}\label{L1}
(AD^2 + BD + C)\Phi = 0,
\end{equation}
where $\Phi$ is a four-element vector defined as 

\begin{equation}\label{L2}
\left\{\hat{u}, \hat{v}, \hat{p}, \hat{T}\right\}^{tr},
\end{equation}
$A$ is represented by the following 4 x 4 matrix,

\begin{equation}\label{L3}
A =
\begin{bmatrix}
1 &0 &0 &0 \\
0 &0 &0 &0 \\
0 &0 &1 &0 \\
0 &0 &0 &1 
\end{bmatrix}
\end{equation}
$B$, $C$ are 4 x 4 matrices as given in appendix I of Malik \cite{Malik}, and $D {\equiv} d/dy$. The boundary conditions associated with the homogeneous ordinary differential equation (\ref{L1}) are given as

\begin{eqnarray}
\phi_1 = \phi_2 = \phi_4 = 0 \hspace{4mm}  in \hspace{4mm} y=0
\end{eqnarray}
\begin{eqnarray}
\phi_1, \phi_2, \phi_4 \rightarrow 0  \hspace{4mm}  as \hspace{4mm} y \rightarrow \infty
\end{eqnarray}
The temperature perturbations become zero at the solid boundary. This assumption is acceptable when the frequency of the disturbance is high. Assuming a temporal stability analysis, equation (\ref{L1}) is discretized using finite difference schemes in the wall normal direction, resulting in an eigenvalue problem in the form

\begin{eqnarray}
E \Phi = \omega F \Phi
\end{eqnarray}
where $E$ and $F$ are 4 x 4 matrices that are obtained from $A$, $B$ and $C$.



\section{Results and Discussion}

\subsection{Preliminaries}

The flow domain consists of a flat-plate boundary layer underlying a hypersonic $M=5.92$ free-stream flow, with the $x$-axis aligned with the wall surface, and the $y$-axis normal to the plate. The length of the computational domain is $600$ mm, and the height of the domain $50$ mm.  The wall deformation height of all wall deformations or the depth of the surface dip is $0.5$ mm, and the deformation is located at $0.5$ m from the leading edge. The Reynolds number based on the boundary layer thickness and the free-stream velocity is $22,750$. The non-dimensional angular frequency of the disturbance is $0.5$, which corresponds to a physical frequency of $132k$ Hz. The wall has a constant temperature equal to $T_{w}$ = 48.69 K, which is equal to the ambient temperature.

Various surface deformations with wall-normal dimensions in the order of the boundary layer displacement thickness are considered in this study: a backward step, a forward step, combination of a backward and a forward step, a combination of a forward and a backward step, a surface hump, a surface dip, a wavy surface with the mean below the wall surface, '{\it sine 1}' (representing successive dips), and a wavy surface with the mean above the wall surface, '{\it sine 2}' (representing successive humps). The computational meshes for all configuration consist of approximately $650,000$ grid points, with appropriate grid resolution at the wall and in the proximity to the wall non-uniformity. The mesh is compressed in the
vicinity of the wall deformation, and stretched to a uniform grid outside of the non-uniformity, such that $\Delta x = 0.1$ mm and $\Delta y_{min} = 0.01$ mm. Spatial coordinates are nondimensionalized by the boundary layer thickness, velocity, density and temperature are scaled by the free-stream velocity, density and temperature, respectively, while the pressure is scaled by the free-stream dynamic pressure.

\begin{figure}[H]
 \begin{center}
    \includegraphics[width=7cm]{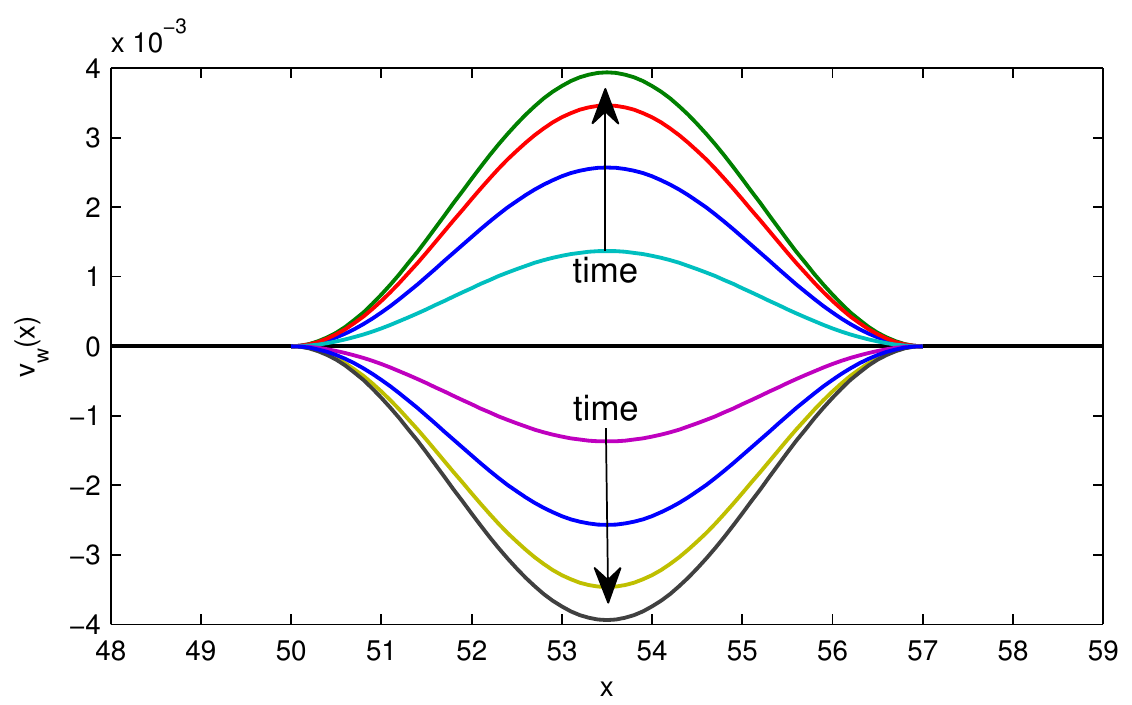}
 \end{center}
  \caption{\label{} Wall disturbance imposed between $x=50$ and $x=57$.}
  \label{f2}
\end{figure}

Two types of wall-normal velocity disturbances are imposed from the wall, in the form

\begin{eqnarray}
v_w(x,t) = A \sin \left[ \pi \frac{(x - x_1)}{(x_2-x_1)} \right]^2 \sin(\omega t),
\end{eqnarray}
where $A$ is the amplitude of the wave, $x_1$ and $x_2$ are the start and the end points along the streamwise direction for the wall disturbance, and $\omega$ is the angular frequency. In this study, $x_1=50$ and $x_2=57$, where the inflow boundary is located in $x=0$; figure \ref{f2} shows the wall-normal velocity disturbance for different time instants. The first disturbance type is a pulse imposed in the time interval $[0,2\pi / \omega]$, which will generate a localized pulse convected with the mean flow in the downstream (the length of the interval correspond to a full period in time). The resulting wave packet will eventually grow or decay as it moves, depending on the initial amplitude and the boundary layer conditions. The second disturbance type is a source of continuous periodic oscillations that are generated inside the boundary layer (for the latter type of disturbance, the wall transpiration disturbance if imposed continuously in the time interval $[0,\infty)$).

\subsection{Grid Convergence Study}

\begin{figure}[H]
 \begin{center}
    \includegraphics[width=16cm]{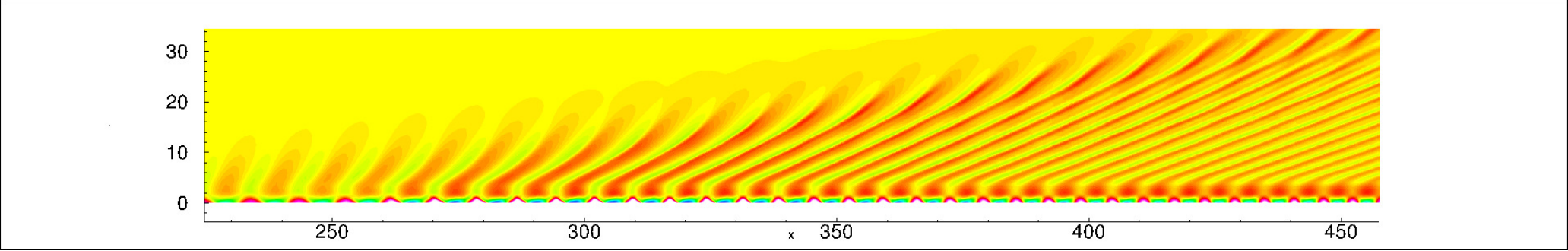}
 \end{center}
  \caption{\label{} Pressure disturbance contours for grid g4.}
  \label{f3}
\end{figure}

This section will examine the effect of the grid resolution on the accuracy of the results for the flat wall case.  The grid densities for each of the 5 cases considered here are listed in Table \ref{t1}.  The coarsest grid had a density of 900 x 180, while the finest grid g5 has a grid density of 2160 x 420. Figure \ref{f3} displays the pressure disturbance contour plot of the g4 grid density case, where the wall disturbance is of periodic continuous blowing-suction type. These contour plots are similar to those reported by Chuvakhov and Fedorov \cite{Chuvakhov}, where an apparent radiation of the disturbance energy to the free-stream is noticed. Chuvakhov and Fedorov revealed that the Mack second mode that is convected inside the boundary layer on a plate with a low wall temperature can radiate acoustic waves into the
free-stream flow. This can be associated with the synchronization of the Mack second mode with slow acoustic waves of
the continuous spectrum.

\begin{table}[H]
\caption{Flat Wall Grid Density Cases}
\centering 
\begin{tabular}{c c c } 
\hline\hline 
Case & Horizontal Density & Vertical Density \\ [0.5ex] 
\hline 
g1 & 900 & 180 \\ 
g2 & 1080 & 240 \\
g3 & 1440 & 300 \\
g4 & 1800 & 360 \\
g5 & 2160 & 420 \\ [1ex] 
\hline 
\end{tabular}
\label{t1} 
\end{table}

\begin{figure}[H]
 \begin{center}
    \includegraphics[width=7cm]{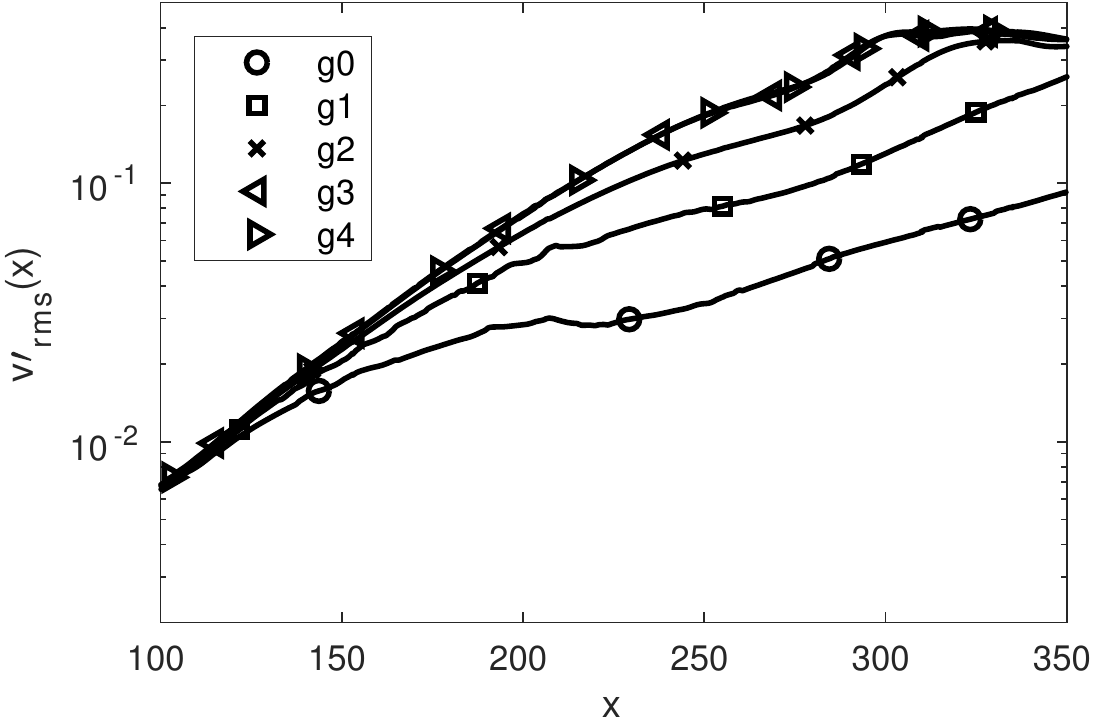}
 \end{center}
  \caption{\label{} Root mean square of wall-normal velocity distribution along the wall ($y=0.7$) for different grid resolutions.}
  \label{f4}
\end{figure}

In figure \ref{f4}, the root mean square of the wall-normal velocity disturbance for the five grid configurations are plotted on a logarithmic scale (in the vertical direction) to determine the convergence of the results. The root mean square was calculated as

\begin{eqnarray}
v'_{rms}(x,y) = \sqrt{\frac{1}{T}\int_{t}^{t+T} [v'(x,y,t)]^2dt},
\end{eqnarray}
where the time span $T$ was in the order of the time it takes a disturbance to go from the inlet boundary to the outlet boundary.

The expected trend with the convergence rate being smaller between grid densities as they become finer is supported by figure \ref{f4}.  The distribution for case g4 is almost identical to case g5, while the difference between cases g1 and g2 is the largest. Based on the results from the grid study (shown in figure \ref{f4}), case g4 (1800 x 360) was selected as the standard grid density for the various wall deformation grids used in the generation of the next results.

\subsection{LST Results}

\begin{figure}[H]
 \begin{center}
   \includegraphics[width=9cm]{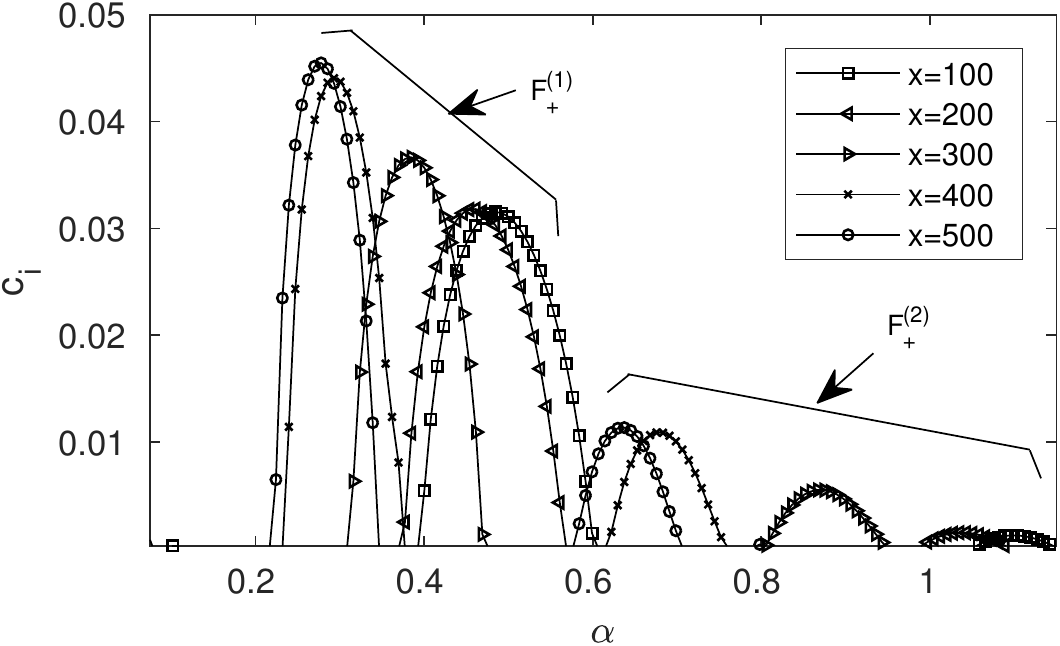}
 \end{center}
  \caption{\label{} Growth rates for different streamwise locations.}
  \label{f5}
\end{figure}

A linear stability analysis is performed to determine the location of the synchronization point, which is where the phase velocities of mode F and mode S become equal. The stability equation was solved for the smooth flat plate with periodic blowing and suction. Typical growth rates are shown in figure \ref{f5} for different streamwise locations: these correspond to Mack's second and third modes, $F_{+}^{1}$ and $F_{+}^{2}$, respectively, according to Fedorov and Tumin's \cite{Fedorov3} terminology. In the cold wall case that is considered in this study, these Mack's higher order modes are associated with fast modes $F$, while for an adiabatic wall they correspond to the slow mode S (Fedorov and Tumin \cite{Fedorov3}). As the streamwise location moves in the downstream, the peak wavenumber of both the second and third Mack modes move upstream, while the amplitudes of both modes increase.

\begin{figure}[H]
 \begin{center}
   \includegraphics[width=10cm]{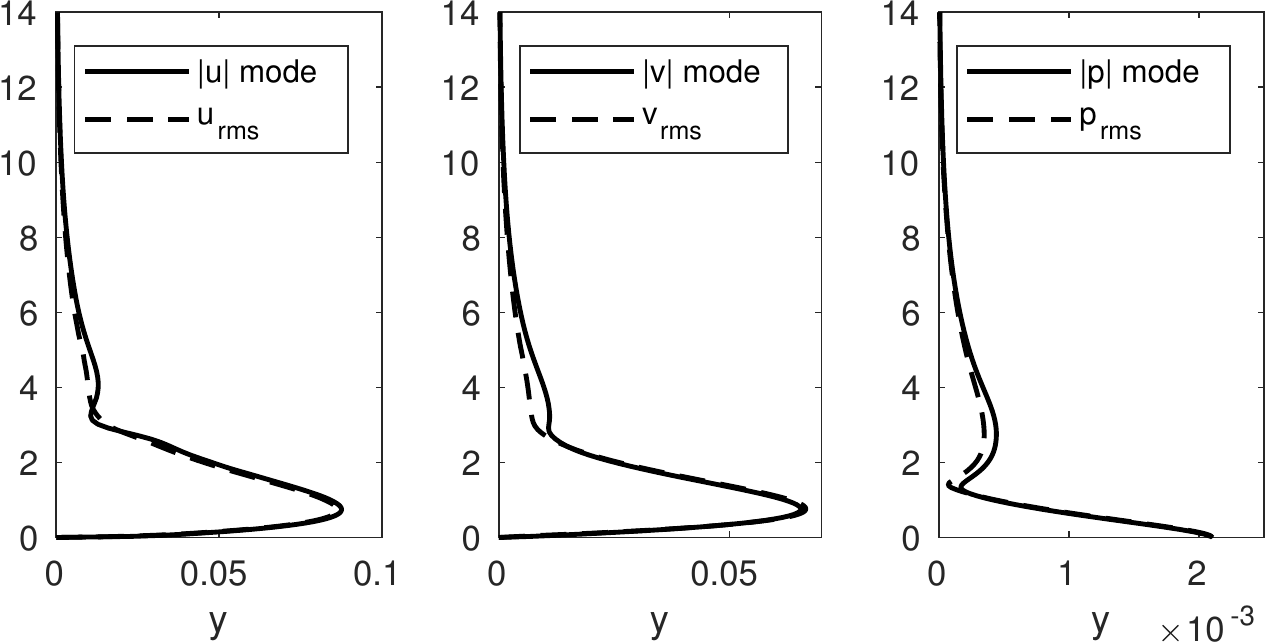}
 \end{center}
  \caption{\label{} Comparison between the modes from linear stability analysis and the root mean square of u-velocity, v-velocity and pressure.}
  \label{f6}
\end{figure}

In figure \ref{f6}, we validate the results from the LST by comparing the absolute values of $u$, $v$ and $p$ first modes with corresponding root mean square quantities (the amplitudes of the LST modes were scaled to match the maximum of the root mean square results). As seen in figure \ref{f6}, the shapes of the curves are in very good agreement, which shows that the linearized stability method is accurate.

\begin{figure}[H]
 \begin{center}
   \includegraphics[width=8cm]{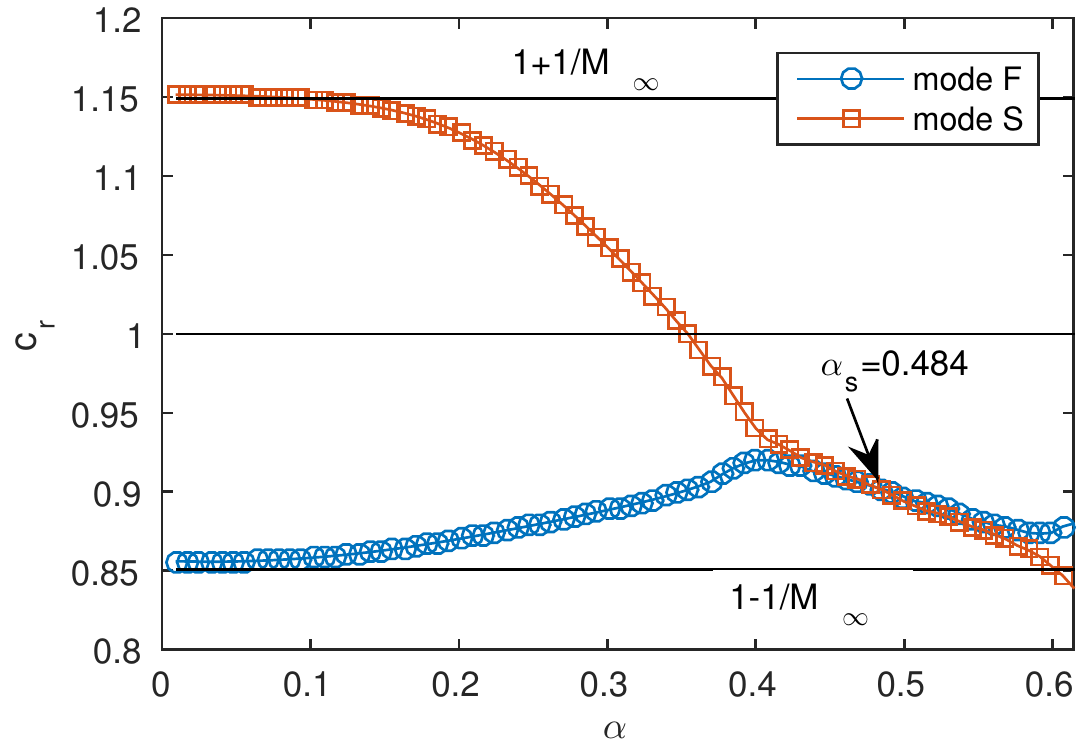}
 \end{center}
  \caption{\label{} Phase velocities for the slow and fast waves as a function of the wavenumber.}
  \label{f7}
\end{figure}

In figure \ref{f7}, the phase speeds of modes F and S are plotted as a function of the wavenumber. At small wavenumbers mode F is moving at $1+1/M_{\infty}$, while mode S is moving at $1-1/M_{\infty}$; as the wavenumber increases the speed of mode F decreases while the speed of mode S increases slightly, until the two become equal where the wavenumber is $\alpha_s$. The synchronization point is calculated using the equation $x_s = (\alpha_s c_r/F)^2/Re)$ (Fong et al. \cite{Fong2}), where $\alpha_s$ is the wavenumber where the two curves intersect each other in figure \ref{f7}, and $F=2\pi f \nu/L_{ref}$ ($f$ is the frequency of the disturbance). For the cases considered in this work the synchronization point was found to be located at $x = 247.6$, which is upstream from the location of wall deformations. As was found in Fong et al. \cite{Fong2}, if the synchronization point is located upstream of the wall roughness then there can potentially be a reduction in the the amplitudes of the traveling waves.

\subsection{Variation of wall deformation shape}

In this section, results from different wall deformation shapes are compared to results from the boundary layer on a flat wall. Wall deformations were described previously, including the grid resolution in the proximity to the deformation (see also figure \ref{f1}). As mentioned previously, all deformations have the same vertical height or depth of $0.5$ mm, which is well inside the boundary layer (it is in the order of the boundary layer displacement thickness, which is equal to $0.583$ mm). The first four deformations in figure \ref{f1} involve sharp edges, while the next four represent smooth non-uniformities in the form of surface humps or dips.

\begin{figure}[H]
 \begin{center}
    \includegraphics[width=8cm]{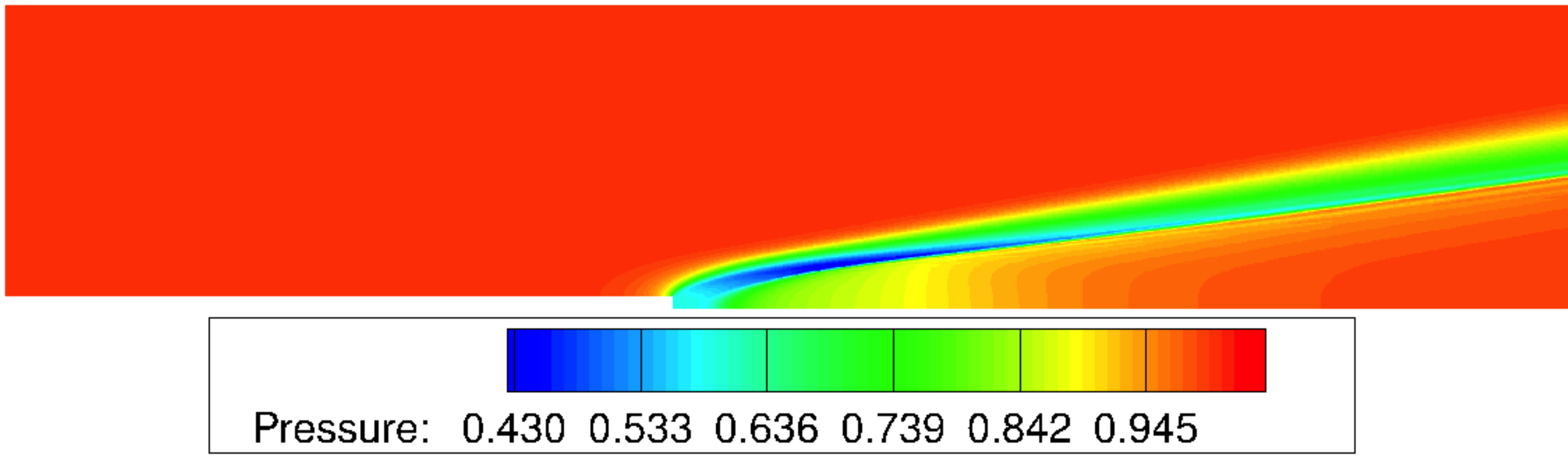}
    \includegraphics[width=8cm]{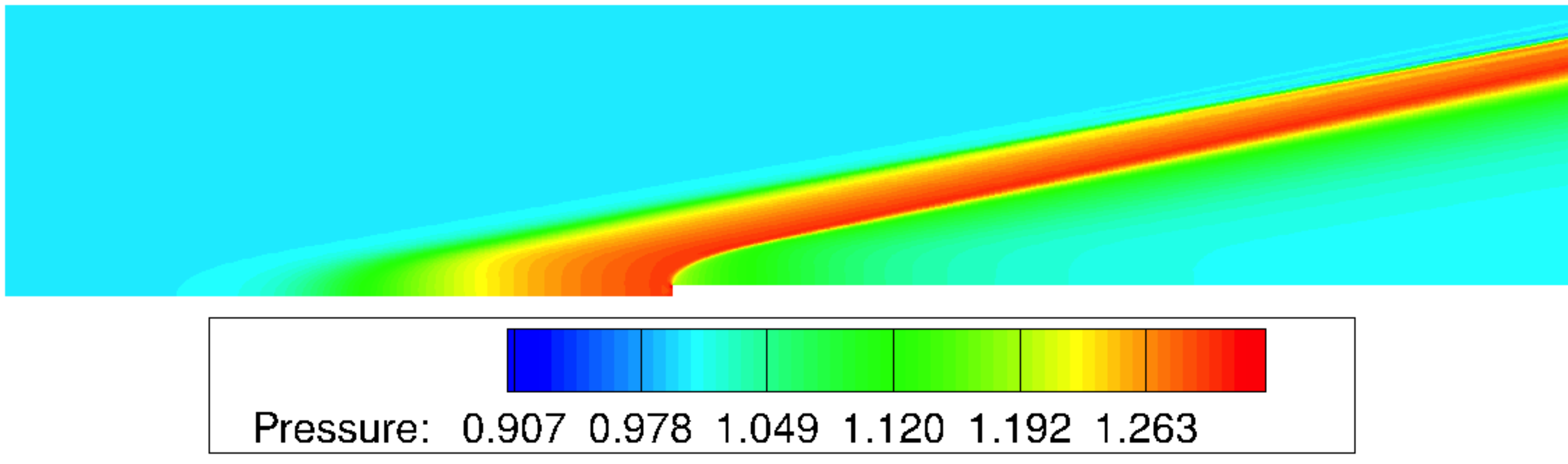}  \\
   a) \hspace{76mm}  b) \\
    \includegraphics[width=8cm]{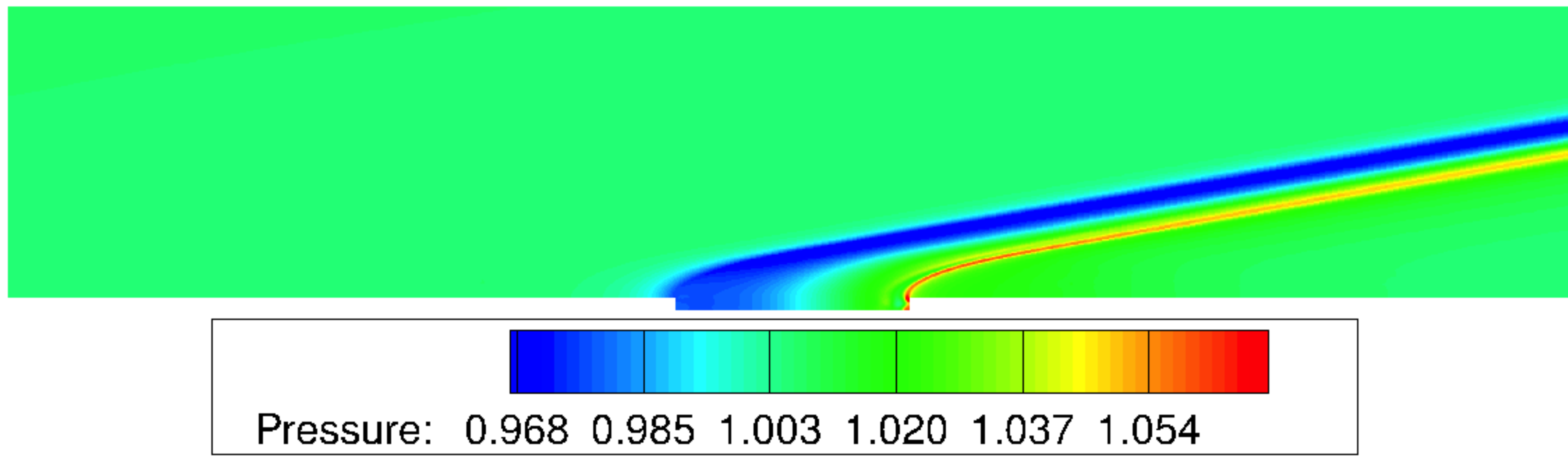}
    \includegraphics[width=8cm]{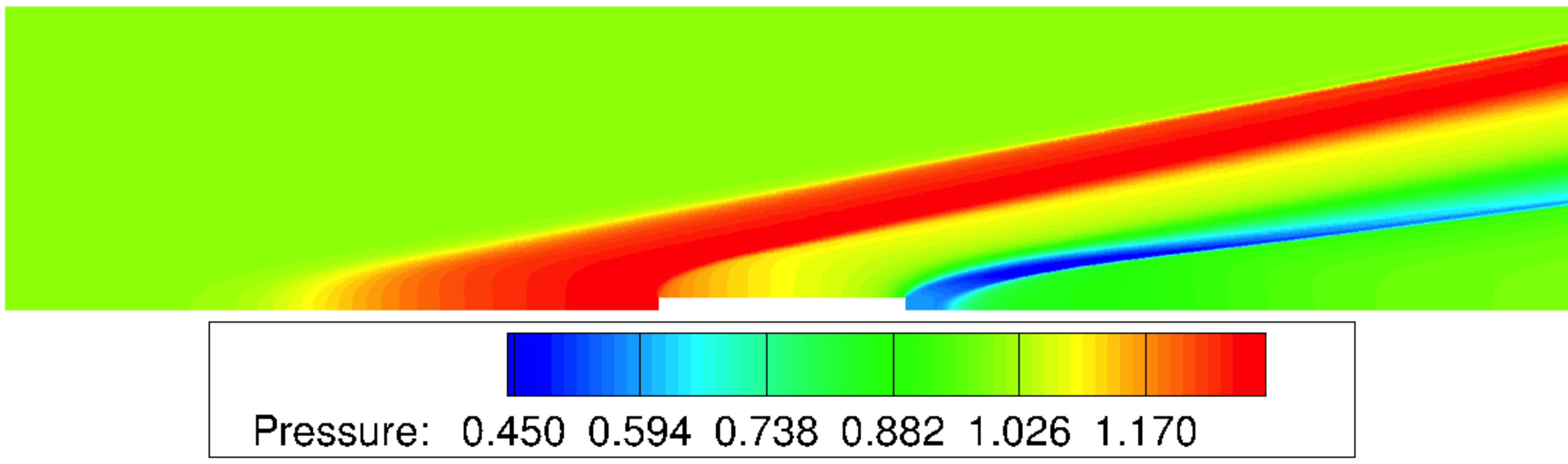}  \\
   c) \hspace{76mm}  d) \\
    \includegraphics[width=8cm]{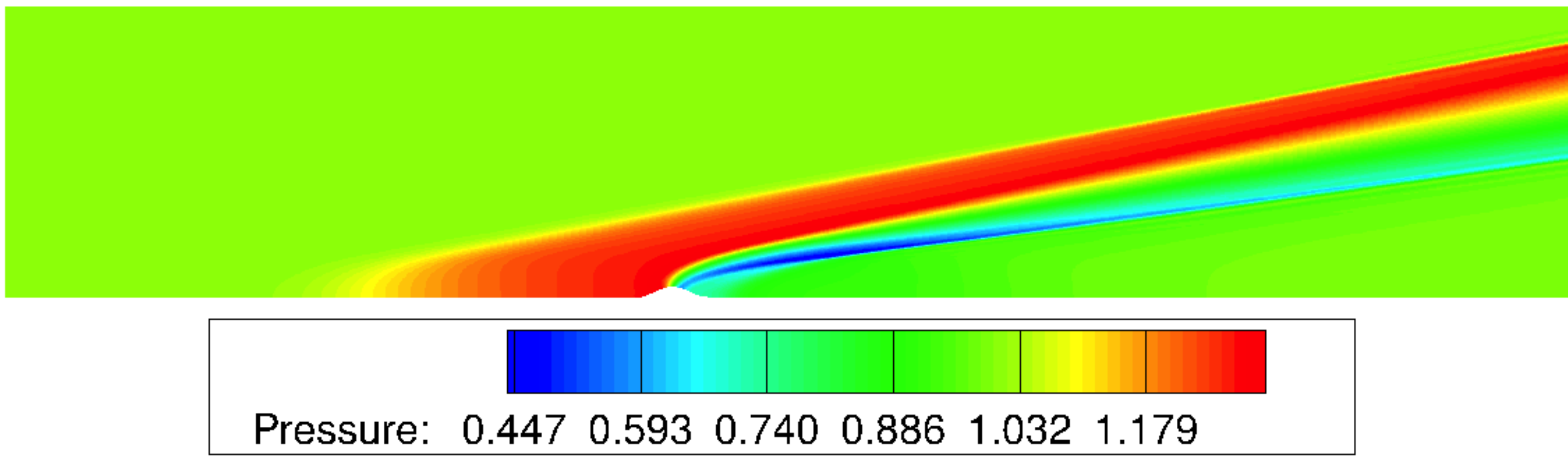}
    \includegraphics[width=8cm]{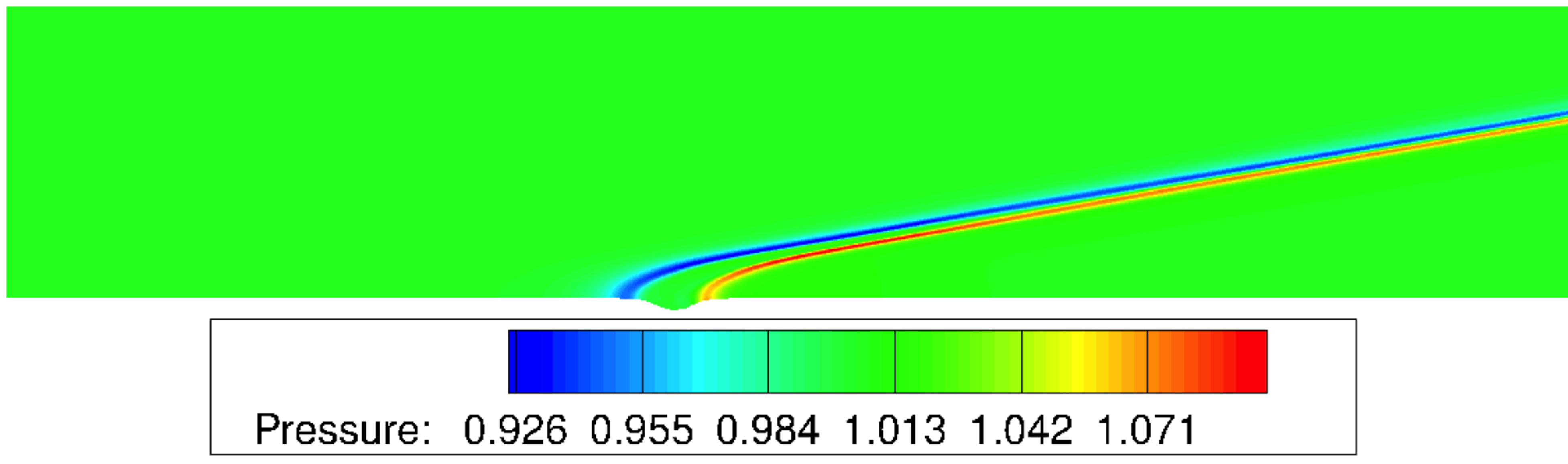}  \\
   e) \hspace{76mm}  f) \\
    \includegraphics[width=8cm]{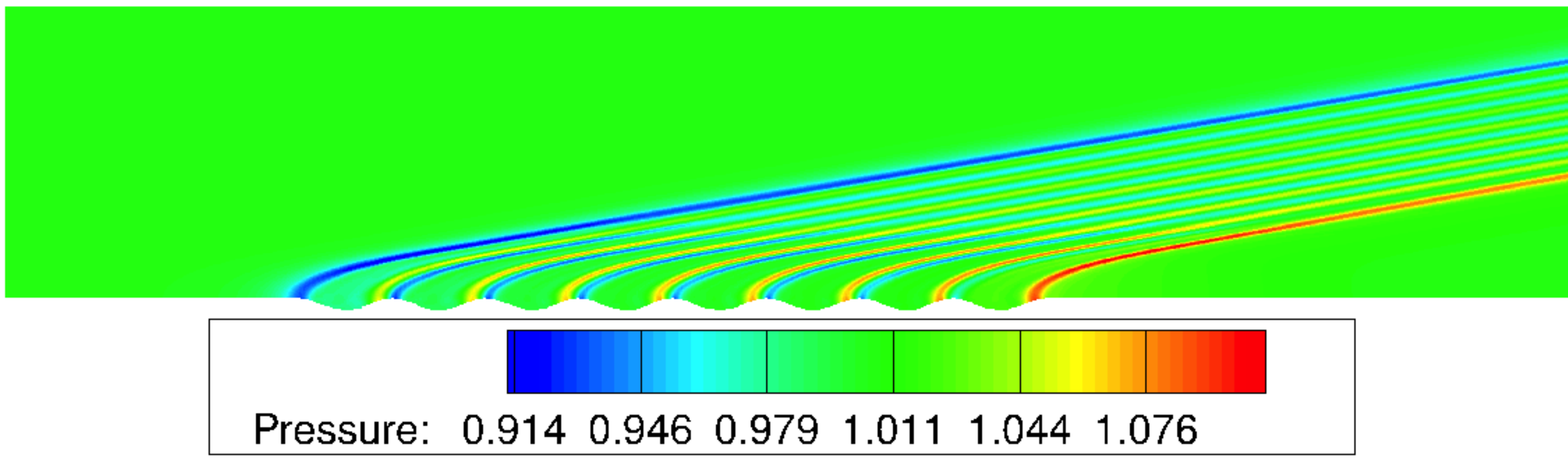}
    \includegraphics[width=8cm]{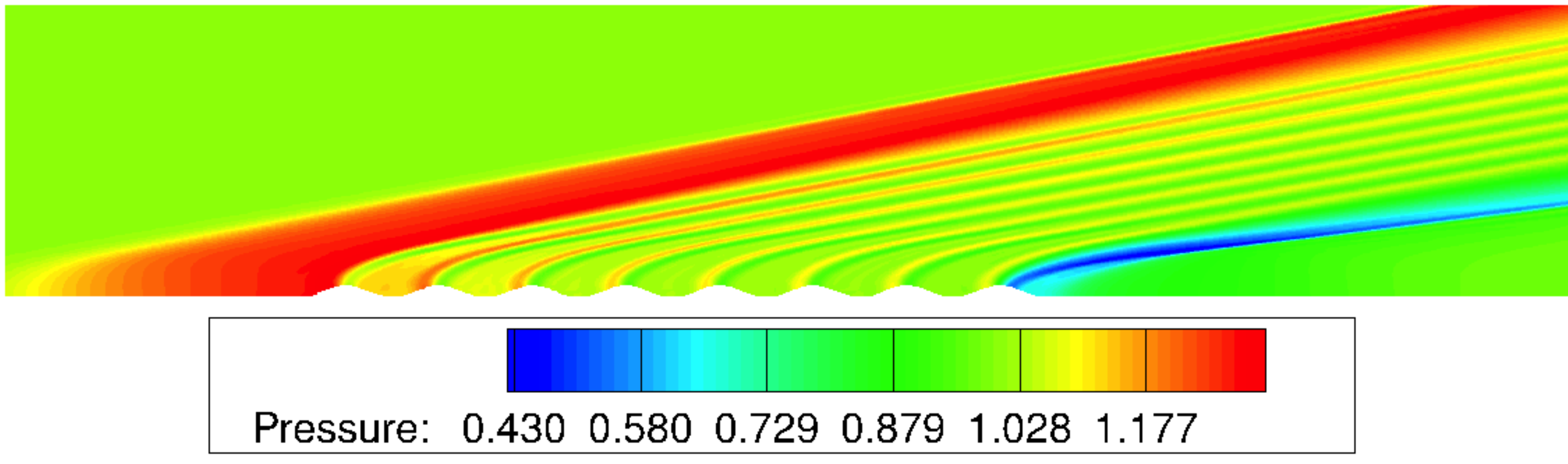}  \\
   g) \hspace{76mm}  h)
 \end{center}
  \caption{\label{} Mean pressure contours in the proximity to the surface deformation: a) backward step; b) forward step; c) combination of a backward and a forward steps; d) combination of a forward and a backward steps; e) hump; f) surface dip; g) sinusoidal shape with the mean below the wall surface (successive dips); h) sinusoidal shape with the mean above the wall surface (successive humps). For all cases, the contour plot frame is: $x_{min}=240$, $x_{max}=400$, $y_{min}=0$, $y_{max}=20$.}
  \label{f8}
\end{figure}

Before the disturbances were imposed from the wall, the mean flow was calculated using inflow profiles corresponding to a compressible boundary layer. Figure \ref{f8} shows contour plots of mean pressure in the vicinity of the backward and forward steps (parts a and b), combinations of backward-forward and forward-backward steps (c and d), the hump (e), the dip (f) and {\it sine 1} and {\it sine 2} (parts g and h, respectively). All types of surface disturbances generated weak discontinuities that extend into the external free-stream flow.  The largest distortion to the mean flow is posed by the hump, {\it sine 2} and the two step deformations that involve a forward step in the upstream. Weak discontinuities are posed by the backward step, the backward-forward step combination, along with the surface dip and the {\it sine 1} deformation.

\begin{figure}[H]
 \begin{center}
    \includegraphics[width=8cm]{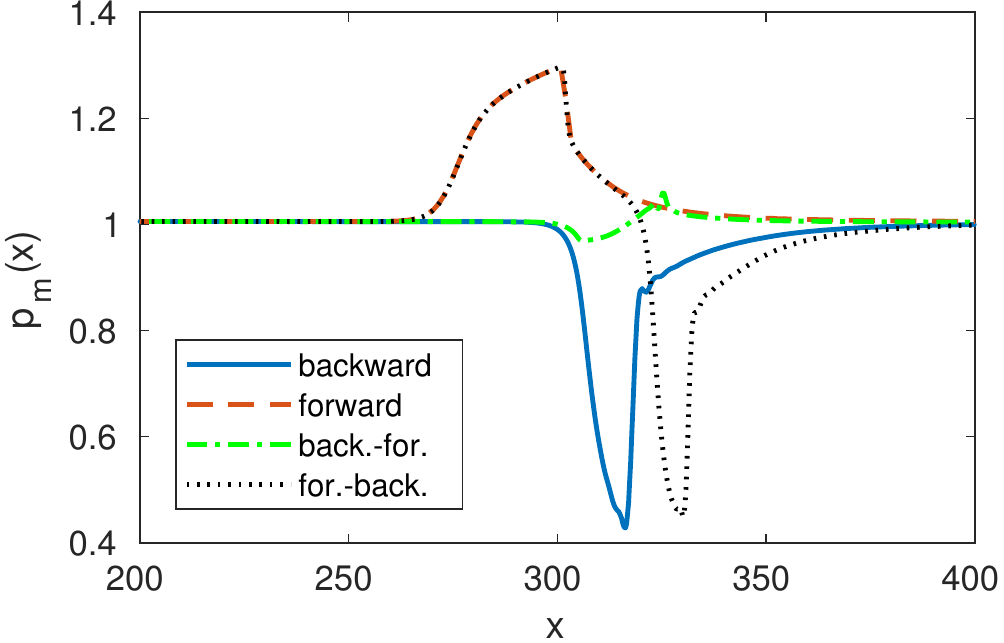}
   \includegraphics[width=8cm]{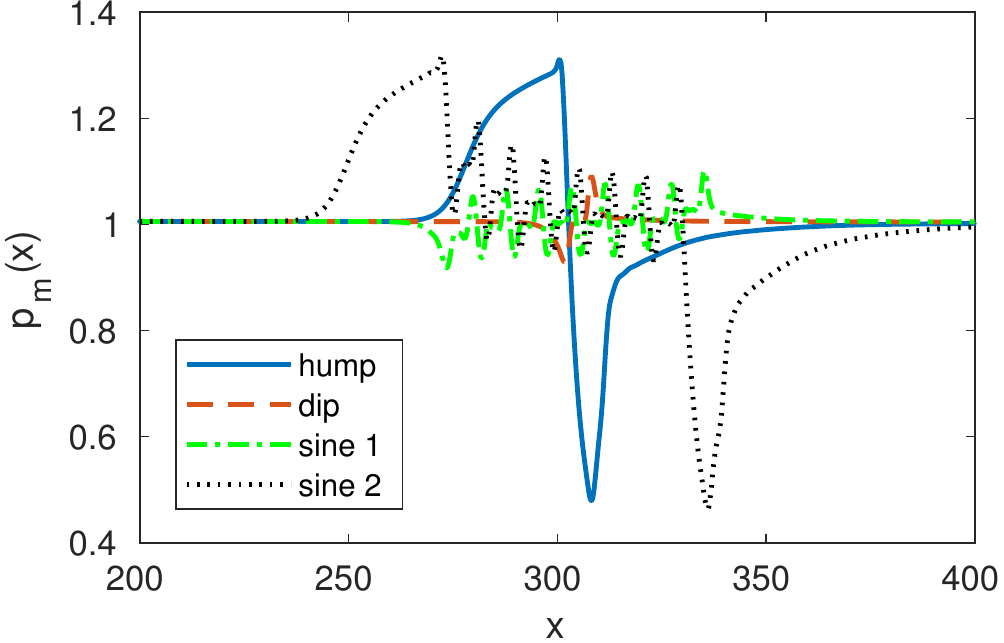}
 \end{center}
  \caption{\label{} Mean pressure distribution along the wall ($y=3$), in the proximity to the surface deformation.}
  \label{f9}
\end{figure}

A quantitative comparison in terms of the mean flow variation along the streamwise direction is revealed in figure \ref{f9}, where mean pressure distributions along the wall are plotted for the eight types of deformations discussed in figure \ref{f8}. The left part of figure \ref{f9} shows only a small distortion of the mean flow by the backward-forward step combination, whereas the backward, forward and forward-backward combination result in larger distortions. Similarly, the right part of figure \ref{f9} illustrates the distortions of the mean flow by the hump, dip and the wavy surface deformations, with the dip and {\it sine 2} resulting in the lowest distortions, as previously observed in the contour plots of figure \ref{f8}. It is interesting to note the balance between the adverse and favorable pressure gradients that are posed by the wall deformations along the streamwise direction, because this will dictate the way the disturbances are affected. Wall deformations that feature a protuberance against the flow, such as the forward step, the hump or the {\it sine 2}, pose an increase in pressure with an associated adverse pressure gradient followed by a decrease in pressure with an associated favorable pressure gradient, or a succession of adverse and favorable pressure gradients. The other wall deformations pose a favorable pressure gradient first followed by an adverse pressure gradient, or a succession of adverse and favorable pressure gradients. Subsequent results will show that the first group of deformations are more likely to reduce the amplitude of disturbances propagating inside the boundary layer.

\begin{figure}[H]
 \begin{center}
    a)  \includegraphics[width=15cm]{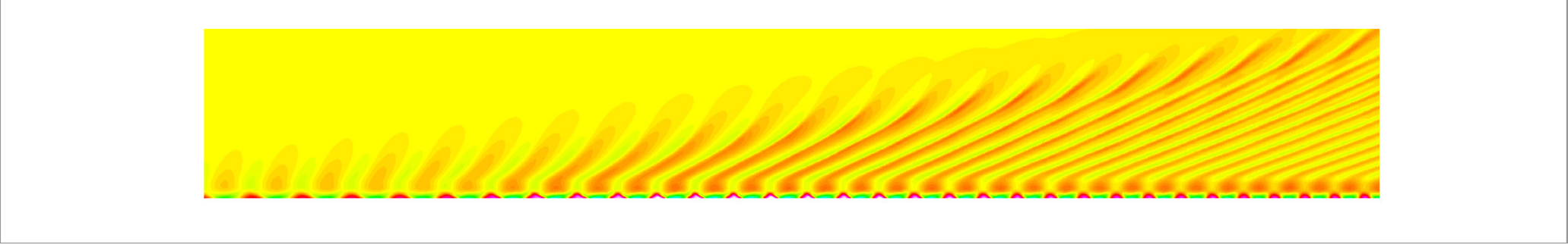} \\
    b)  \includegraphics[width=15cm]{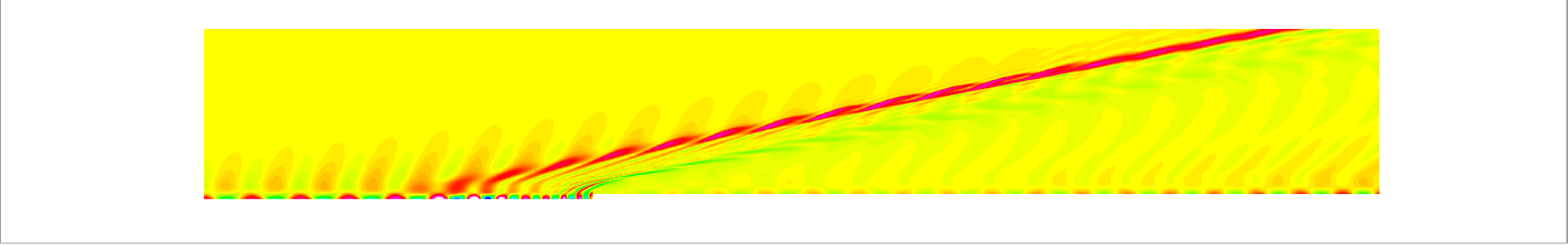} \\
   c)   \includegraphics[width=15cm]{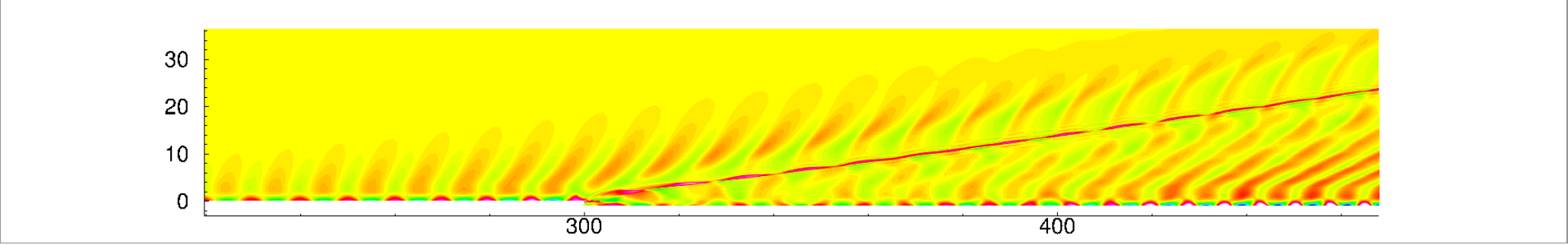}
 \end{center}
  \caption{\label{} Pressure disturbance contours in the proximity to: a) flat wall; a) forward step; b) combination of a backward step.}
  \label{f10}
\end{figure}

\begin{figure}[H]
 \begin{center}
  a)  \includegraphics[width=13cm]{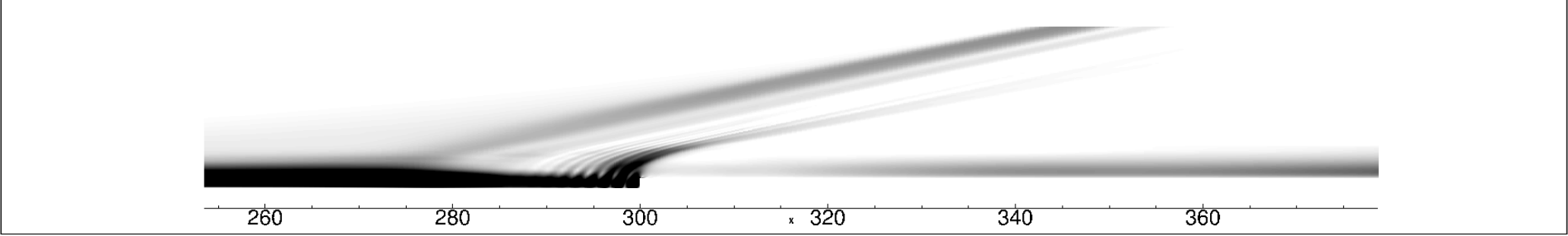} \\
  b)  \includegraphics[width=13cm]{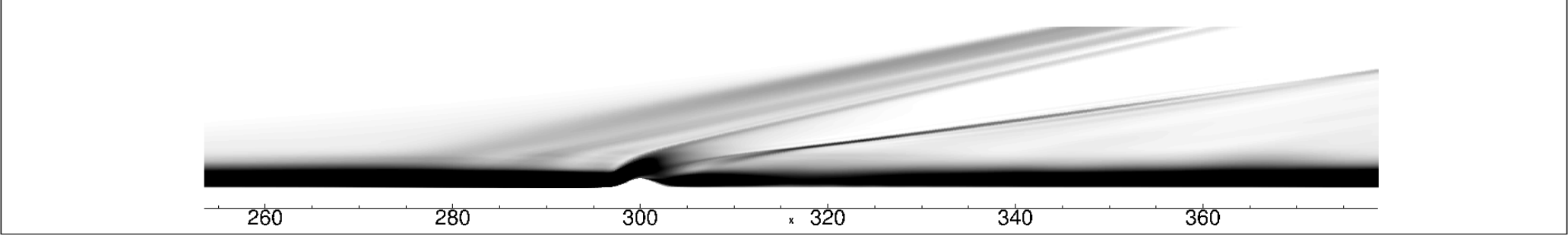} \\
  c)  \includegraphics[width=13cm]{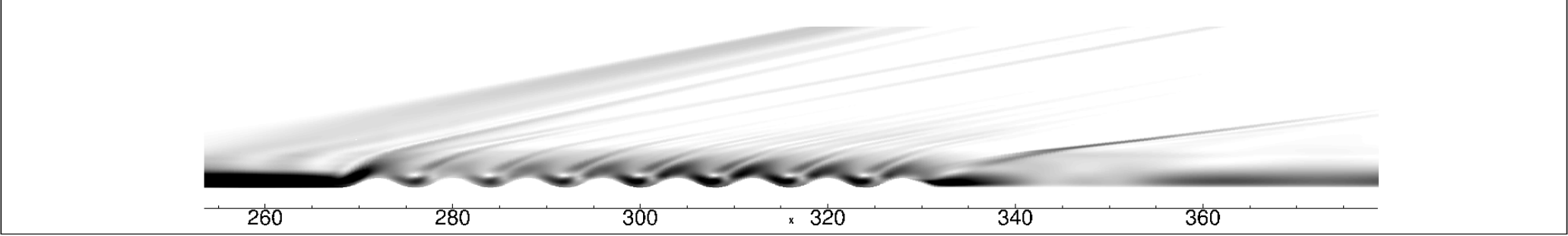}
 \end{center}
  \caption{\label{} Time-averaged kinetic energy contours in the proximity to the surface deformations: a) forward step; b) hump; c) wavy surface type 2. The ratio between 'gray' and 'black' patches is approximately $1/10$}.
  \label{f11}
\end{figure}

Contour plots of pressure disturbance in the proximity to the location of the surface non-uniformity are given in figure \ref{f10} for the flat wall and two types of wall deformations, one that presented the greatest distortion to the mean flow (forward step), and one that presented the smallest distortion (combination of backward and forward steps) in figure \ref{f8}. The contour plots correspond to the periodic blowing and suction disturbance case. In this figure, one can notice that for the forward step case the disturbances are deviated and forced to follow the discontinuity line that extends outside the boundary layer. This deviation of energy to the external flow could partially explain the reduction of disturbance energy inside the boundary layer. To support this presumption, in figure \ref{f11} contours of the time-averaged kinetic energy in the proximity to the wall deformations is plotted for the most effective roughness elements (forward step, hump, and the wavy surface). This figure suggests that a portion of the kinetic energy is directed to the external flow (this does not happen in subsonic boundary layers because there are no discontinuities posed by the roughness element). However, the deviated portion is small with respect to the upstream kinetic energy level (roughly $10\%$), so this may not be the main mechanism of disturbance energy reduction, but definitely something that should be taken into account). For the other case shown in figure \ref{f10}c, the deviation is not so significant since the discontinuity is weak and there is a favorable pressure gradient in the upstream, as revealed by figures \ref{f8} or \ref{f9}. While these two cases are the extremes, the other cases behave similarly (not shown here), depending on whether the pressure gradient is adverse or favorable, or depending on the intensity of the discontinuity.

\begin{figure}[H]
 \begin{center}
   \includegraphics[width=6cm]{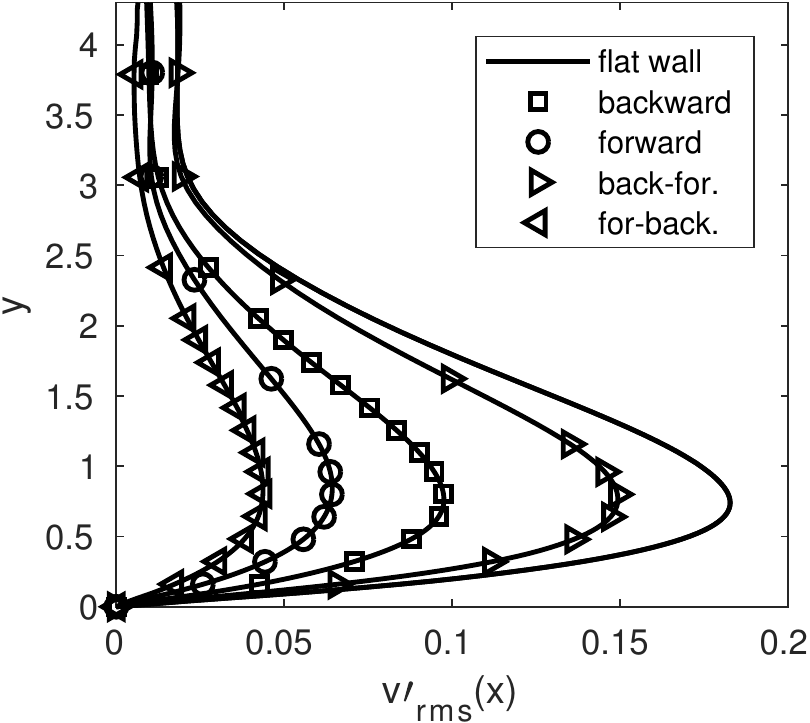}
   \includegraphics[width=6cm]{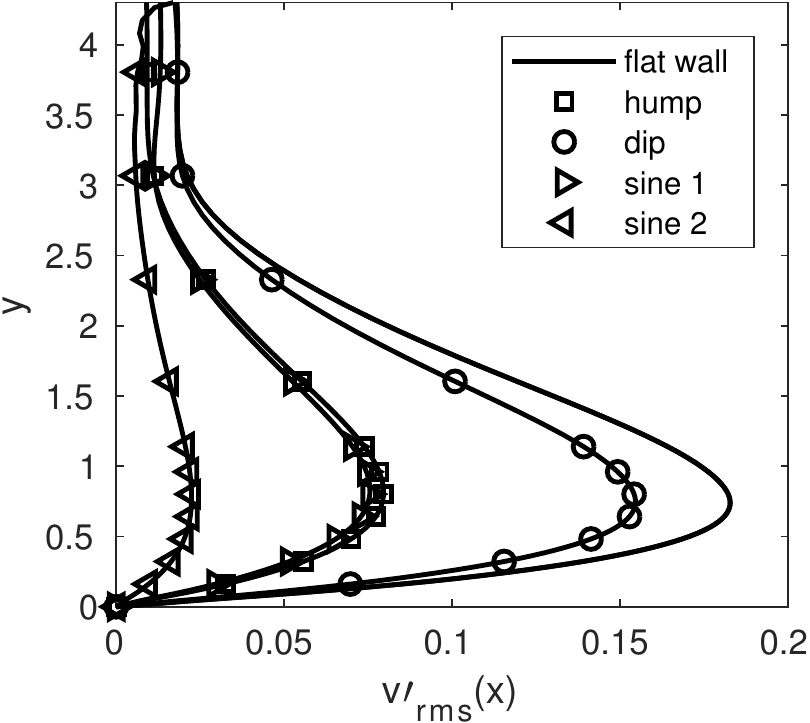}
 \end{center}
  \caption{\label{} Root mean square of wall-normal velocity distribution along the wall ($y=0.7$) for the pulse disturbance.}
  \label{f12}
\end{figure}

\begin{figure}[H]
 \begin{center}
   \includegraphics[width=6cm]{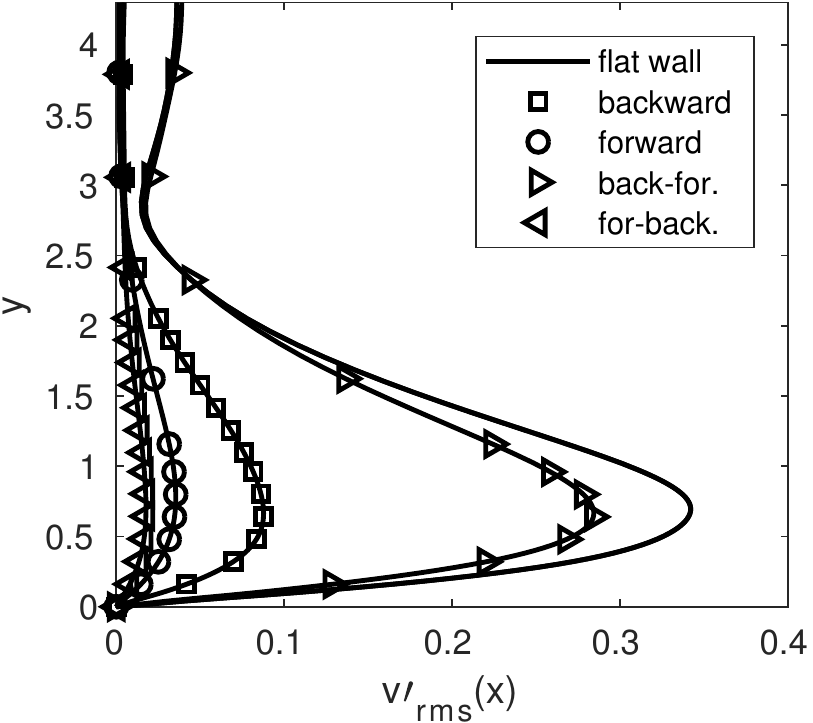}
   \includegraphics[width=6cm]{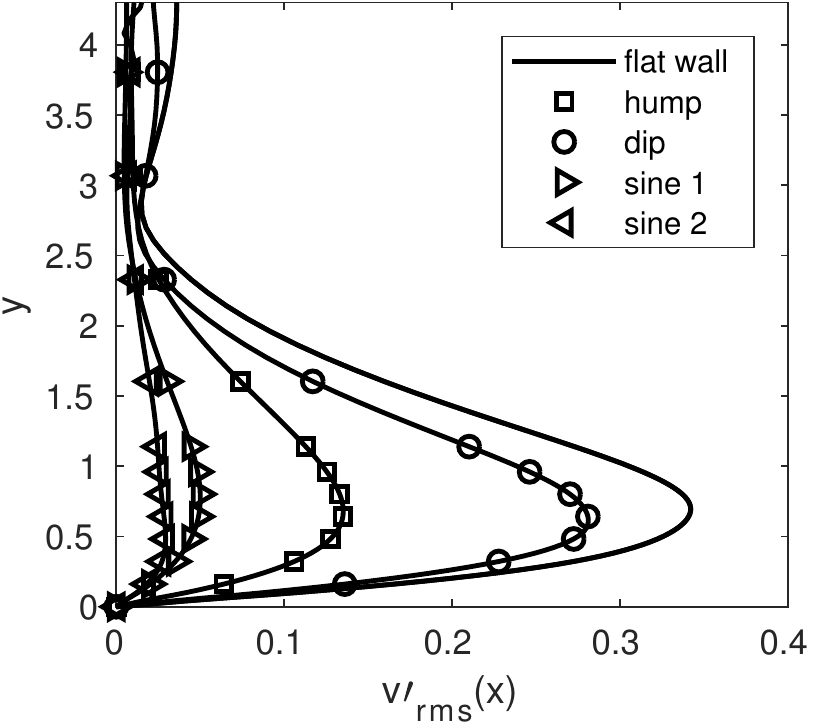}
 \end{center}
  \caption{\label{} Root mean square of wall-normal velocity distribution along the wall ($y=0.7$) for the periodic blowing and suction disturbance.}
  \label{f13}
\end{figure}

A quantitative measure of energy reduction is shown in the next figures, where the root mean square distributions of wall-normal velocity are plotted as a function of the streamwise or wall-normal directions. In figures \ref{f12} and \ref{f13}, the root mean square of wall-normal velocity are plotted for the pulse and periodic blowing and suction disturbances, respectively, and compared to the result from the flat surface. For the pulse disturbance, {\it sine 2} and the forward-backward step combination reduce the disturbance amplitude the most. In the case of the periodic blowing/suction, both 
sinusoidal deformations and the forward and forward-backward steps are sufficiently reducing the wall-normal velocity disturbance amplitudes. The effect from the wall dip seems to be negligible for both disturbances, while the hump has moderate
effectiveness, mainly due to its localized character, especially when compared to the extended length deformations.

\begin{figure}[H]
 \begin{center}
   \includegraphics[width=6cm]{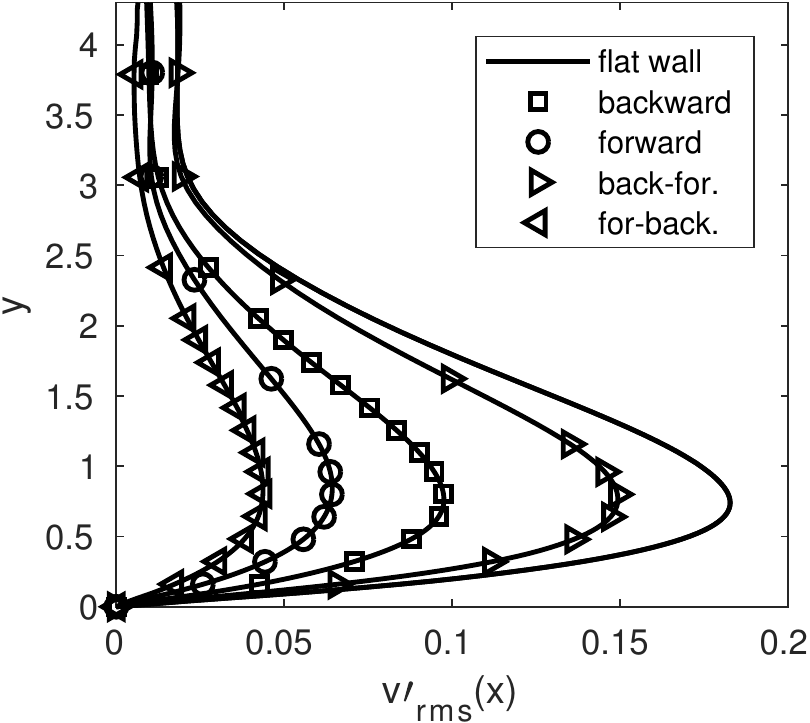}
   \includegraphics[width=6cm]{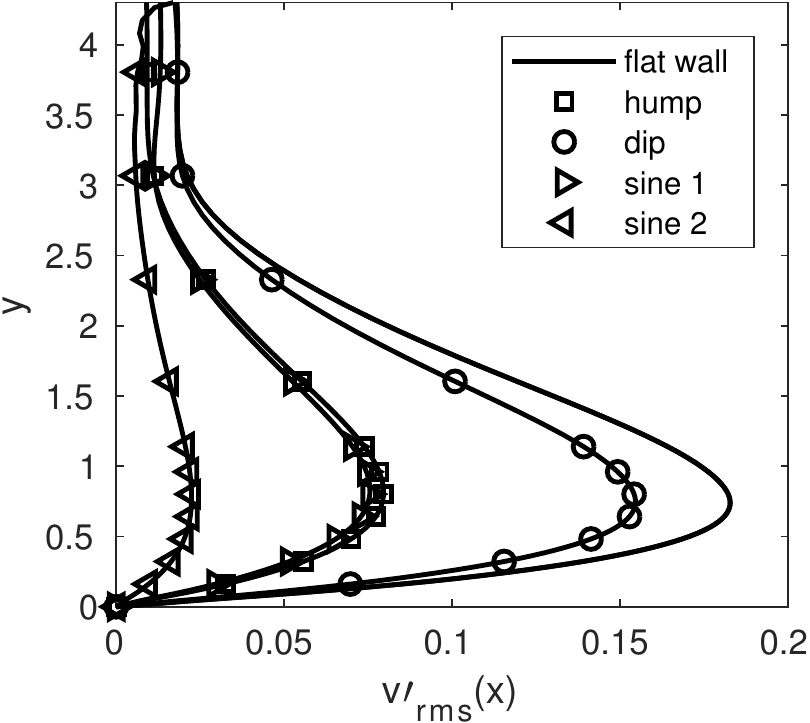}
 \end{center}
  \caption{\label{} Profiles of root mean square of wall-normal velocity ($x=350$) for the pulse disturbance.}
  \label{f14}
\end{figure}

\begin{figure}[H]
 \begin{center}
   \includegraphics[width=6cm]{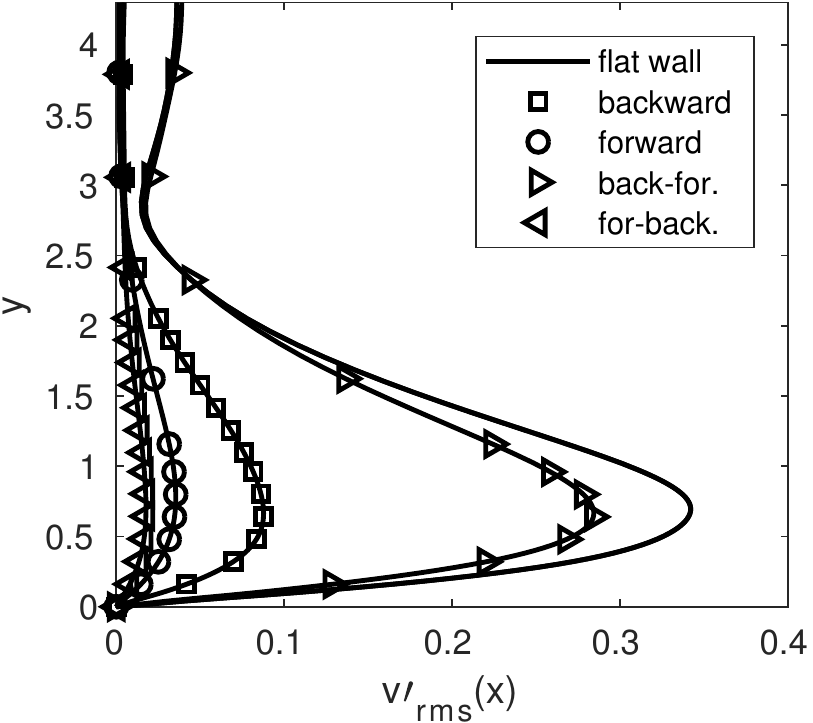}
   \includegraphics[width=6cm]{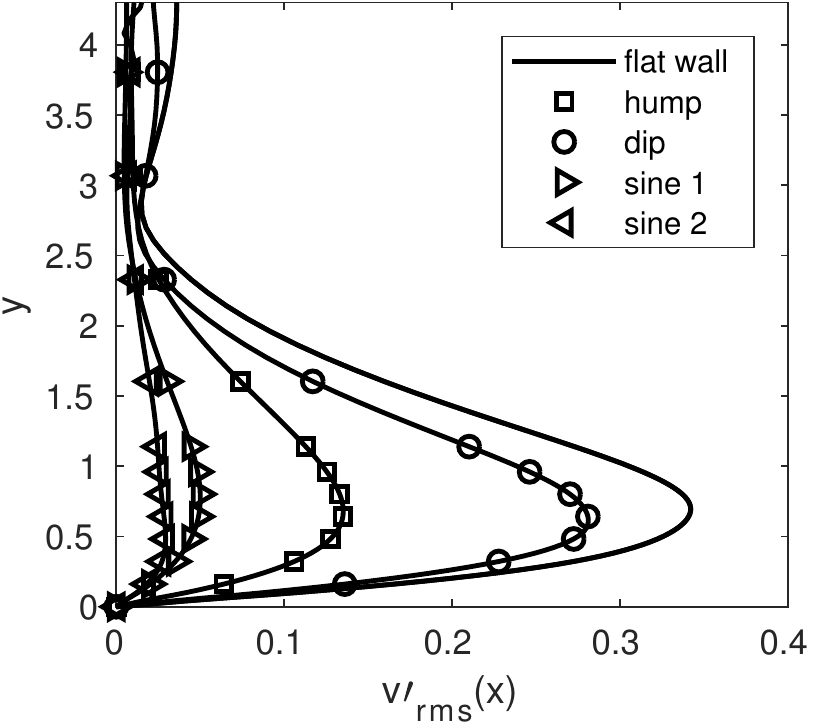}
 \end{center}
  \caption{\label{} Profiles of root mean square of wall-normal velocity ($x=350$) for the periodic blowing and suction disturbance.}
  \label{f15}
\end{figure}

A set of results concerning the distribution of the root mean square wall-normal velocity in the vertical direction are given in figures \ref{f14} and \ref{f15}, for $x=350$. The profiles of wall-normal velocity suggest that most of the disturbance kinetic energy is confined inside the boundary layer (with thickness in the order of $3.5$ deformation heights). The {\it sine 2} is the most effective in the case of the pulse disturbance, with the forward-backward step combination having the greatest effects in the case of the blowing/suction disturbance. In general, all types of deformations
are capable of reducing the amplitude of the disturbances to some extent based on their shape and localized behavior, except for the dip and backward-forward step combination, which have less significant effects.

\subsection{Variation of the wall deformation width}

\begin{figure}[H]
 \begin{center}
    \includegraphics[width=13cm]{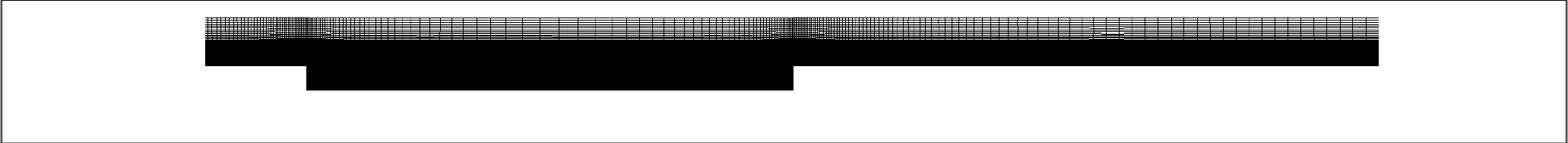} \\
    \includegraphics[width=13cm]{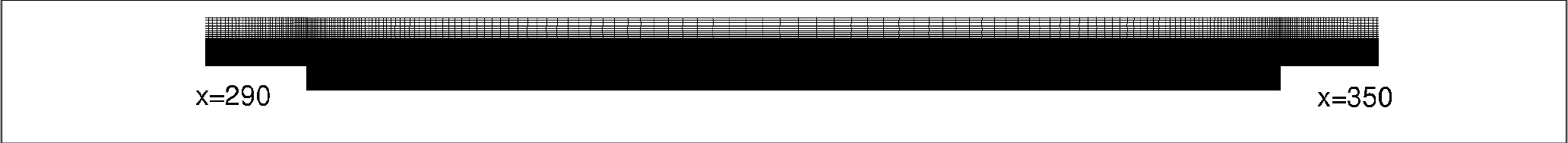}
 \end{center}
  \caption{\label{} Backward-forward step shapes.}
  \label{f16}
\end{figure}

\begin{figure}[H]
 \begin{center}
    \includegraphics[width=6cm]{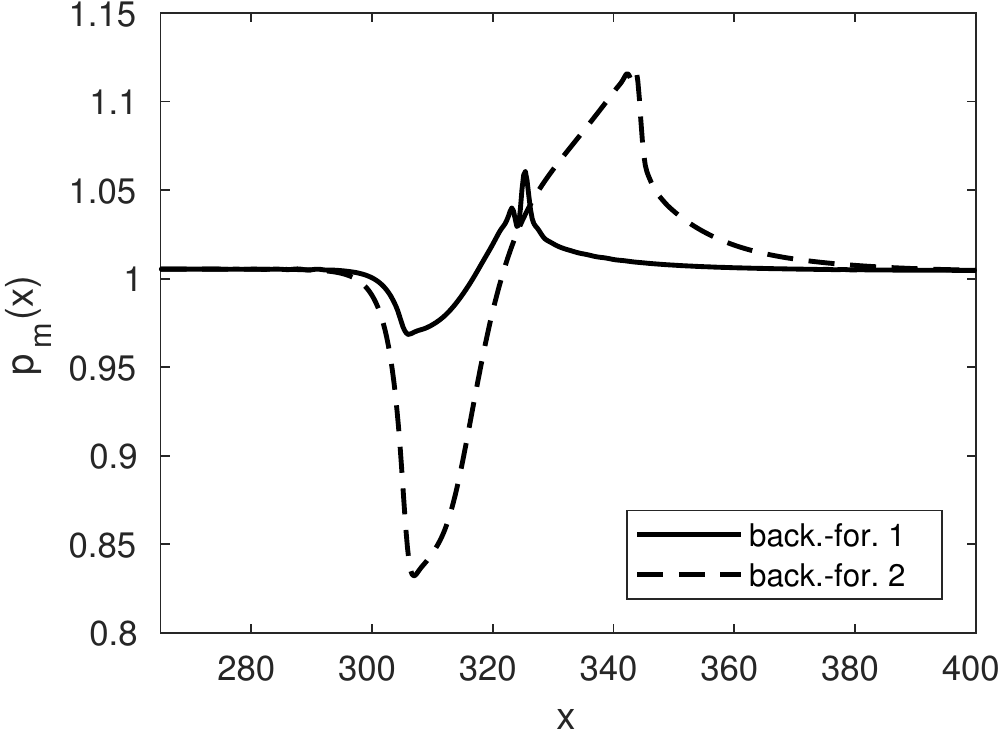}
   \includegraphics[width=6cm]{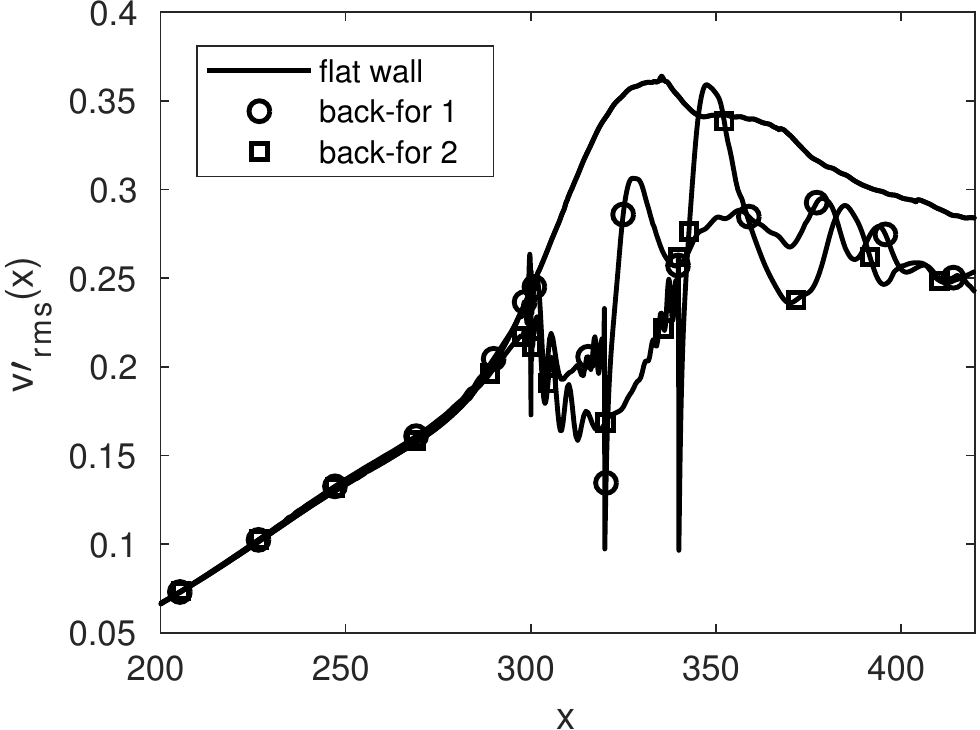} \\
   \hspace{10mm}  a) \hspace{70mm}  b)
 \end{center}
  \caption{\label{} a) Mean pressure distribution along the wall ($y=1$); b) Root mean square of wall-normal velocity distribution along the wall ($y=0.7$).}
  \label{f17}
\end{figure}

In this section, the dependency of the energy reduction on the streamwise width of the deformation is investigated for the backward-forward step combination, forward-backward combination, the hump, the dip, and the wavy surfaces of both types. 
Figure \ref{f16} shows the two backward-forward step configurations that are considered: one has the length of $20$ and the other $40$ step heights. In figure \ref{f17}a, the mean pressure distribution along the wall reveals that the extension of the width between the backward and the forward steps poses a greater distortion in the mean flow, compared to its smaller width. This is because the boundary layer flow in the second case (higher width) has enough room to adjust itself to the original upstream condition, so the interaction with the forward step becomes stronger; in other words, the two flows in the vicinity of the wall deformations are less affected by each other as the width is increased. However, in figure \ref{f17}b one can notice that the two deformations have almost the same effect on the propagating disturbance.

\begin{figure}[H]
 \begin{center}
    \includegraphics[width=13cm]{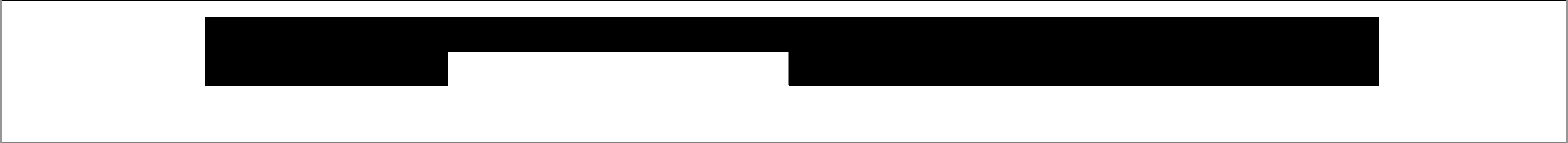} \\
    \includegraphics[width=13cm]{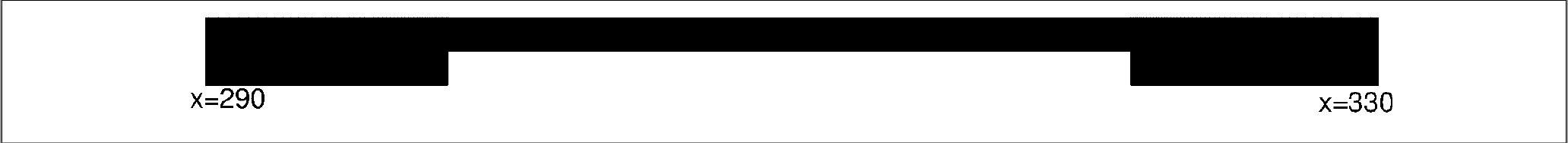}
 \end{center}
  \caption{\label{} Forward-backward step shapes.}
  \label{f18}
\end{figure}

\begin{figure}[H]
 \begin{center}
    \includegraphics[width=6cm]{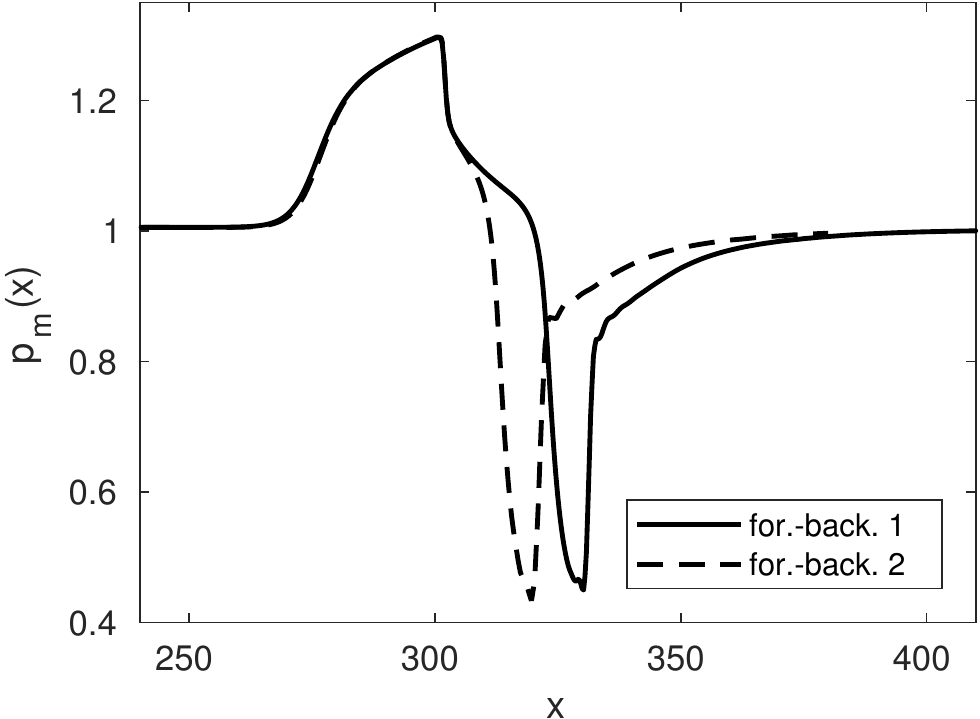}
   \includegraphics[width=6cm]{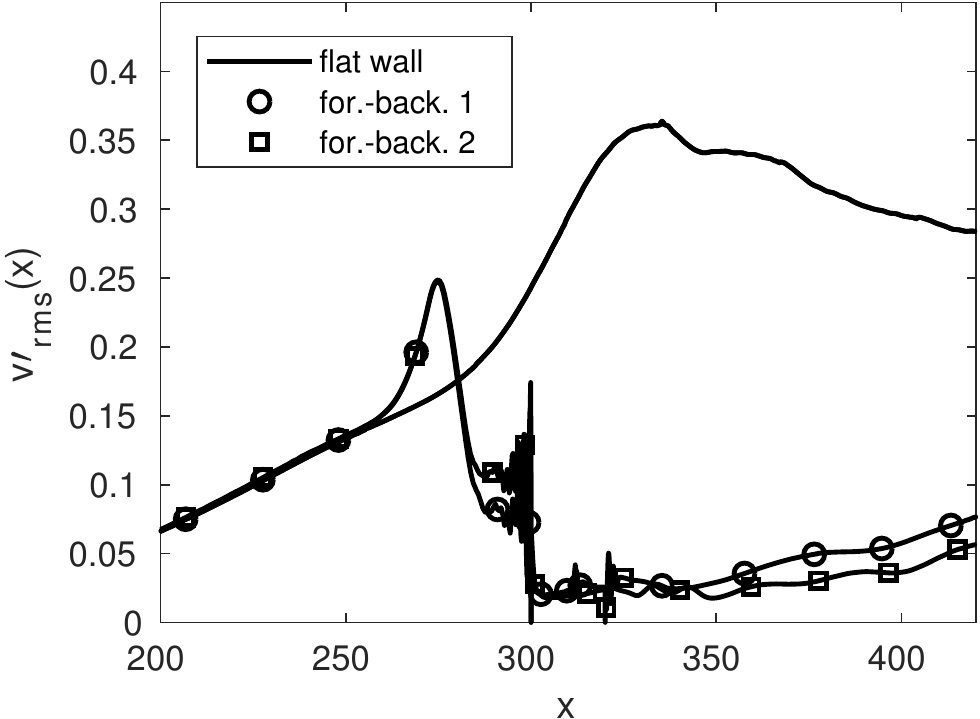} \\
   \hspace{10mm}  a) \hspace{70mm}  b)
 \end{center}
  \caption{\label{} a) Mean pressure distribution along the wall ($y=1$); b) Root mean square of wall-normal velocity distribution along the wall ($y=0.7$).}
  \label{f19}
\end{figure}

The same analysis is performed for the forward-backward combination, as shown in figure \ref{f18}, where the top part corresponds to a width of $12$ step heights, while the bottom to $20$ step heights. The upstream adverse pressure gradient in figure \ref{f19}a does not reveal any difference between the two cases, while there is some difference in the favorable pressure gradient portion in the downstream. Anyway, the effect on the disturbance propagation is almost the same, as displayed by distributions of root mean square of wall-normal velocity disturbance in figure \ref{f19}b.

\begin{figure}[H]
 \begin{center}
    \includegraphics[width=13cm]{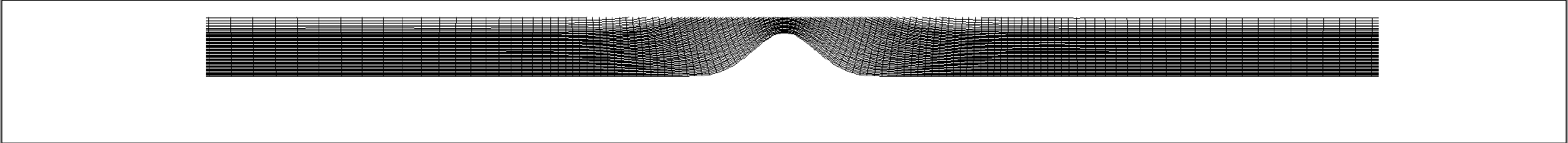} \\
    \includegraphics[width=13cm]{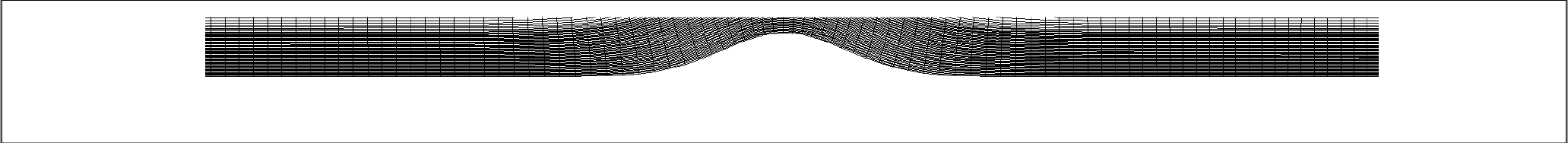} \\
   \includegraphics[width=13cm]{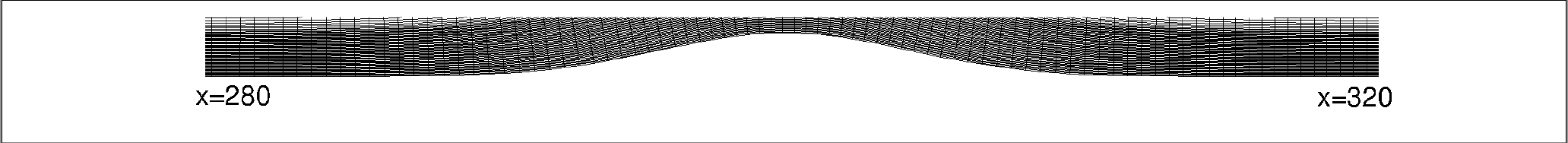}
 \end{center}
  \caption{\label{} Hump shapes.}
  \label{f20}
\end{figure}

\begin{figure}[H]
 \begin{center}
    \includegraphics[width=6cm]{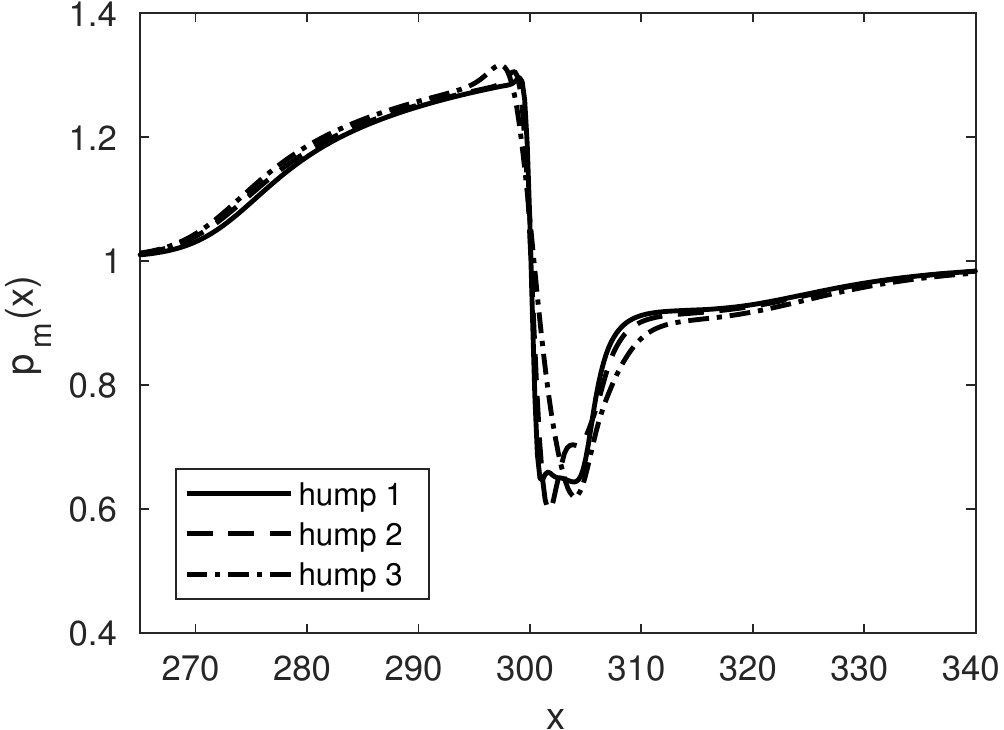}
    \includegraphics[width=6cm]{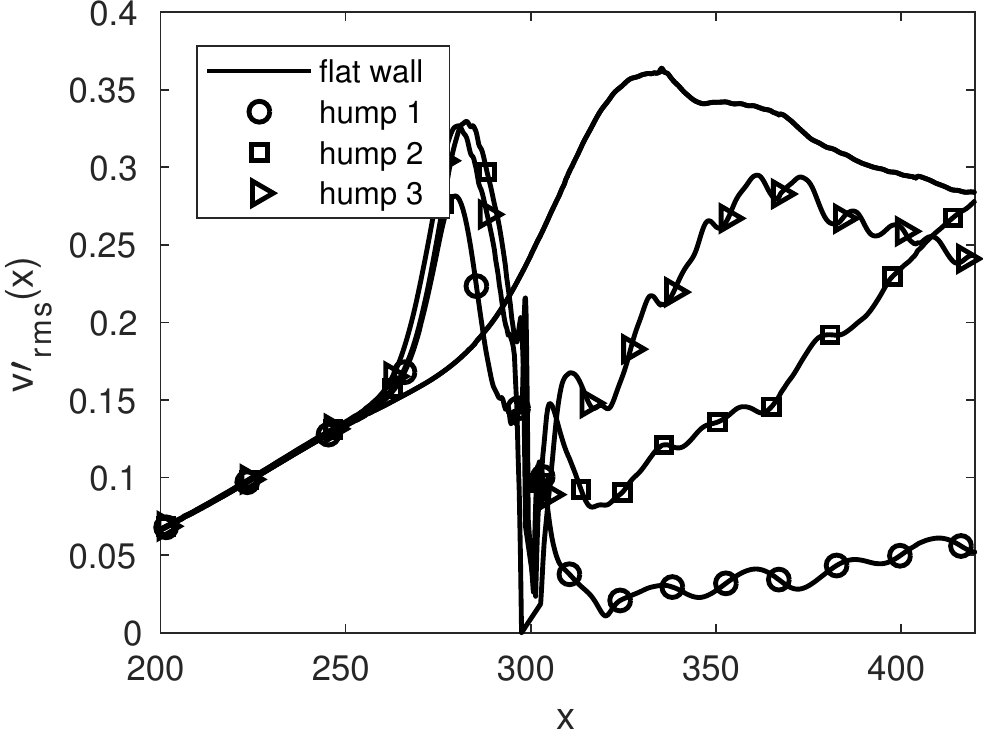} \\
   \hspace{10mm} a) \hspace{78mm} b)
 \end{center}
  \caption{\label{} a) Mean pressure distribution along the wall ($y=1$); b) Root mean square of wall-normal velocity distribution along the wall ($y=0.7$).}
  \label{f21}
\end{figure}

Increasing the streamwise width of the hump (see figure \ref{f20}) has an effect in the mean pressure distribution as shown in figure \ref{f21}a, but not significant. As far as the root mean square of wall-normal velocity is concerned (see figure \ref{f21}b), an increase of the width of the hump deformation results in
less significant reduction of the disturbance amplitudes leading to the conclusion that a shorter, more localized hump should be utilized to reduce the disturbance energy. This is somehow in contrast to what Fong et al. \cite{Fong3} found, but it must be mentioned that the roughness considered in that study was not smooth (at the matching between the roughness element and the wall), and that the disturbance was a pulse propagating as a wave packet, while here the disturbance is periodic.

\begin{figure}[H]
 \begin{center}
    \includegraphics[width=13cm]{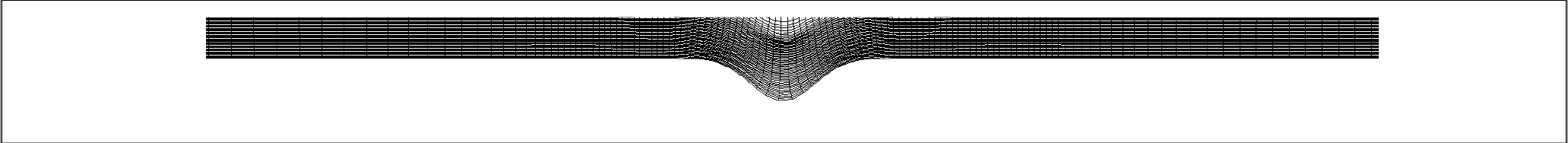} \\
    \includegraphics[width=13cm]{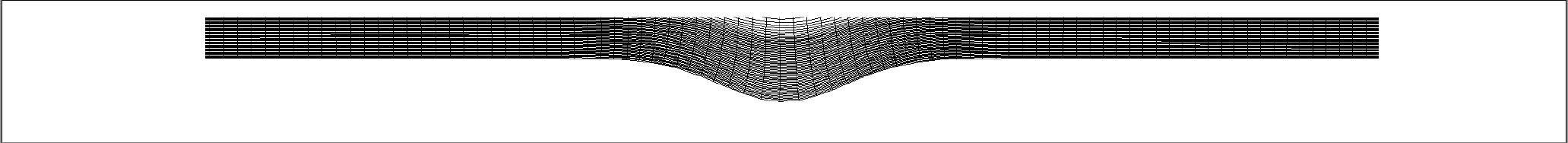} \\
   \includegraphics[width=13cm]{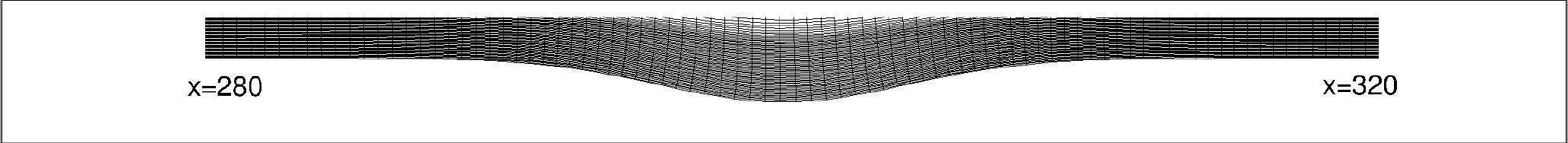}
 \end{center}
  \caption{\label{} Dip shapes.}
  \label{f22}
\end{figure}

\begin{figure}[H]
 \begin{center}
    \includegraphics[width=6cm]{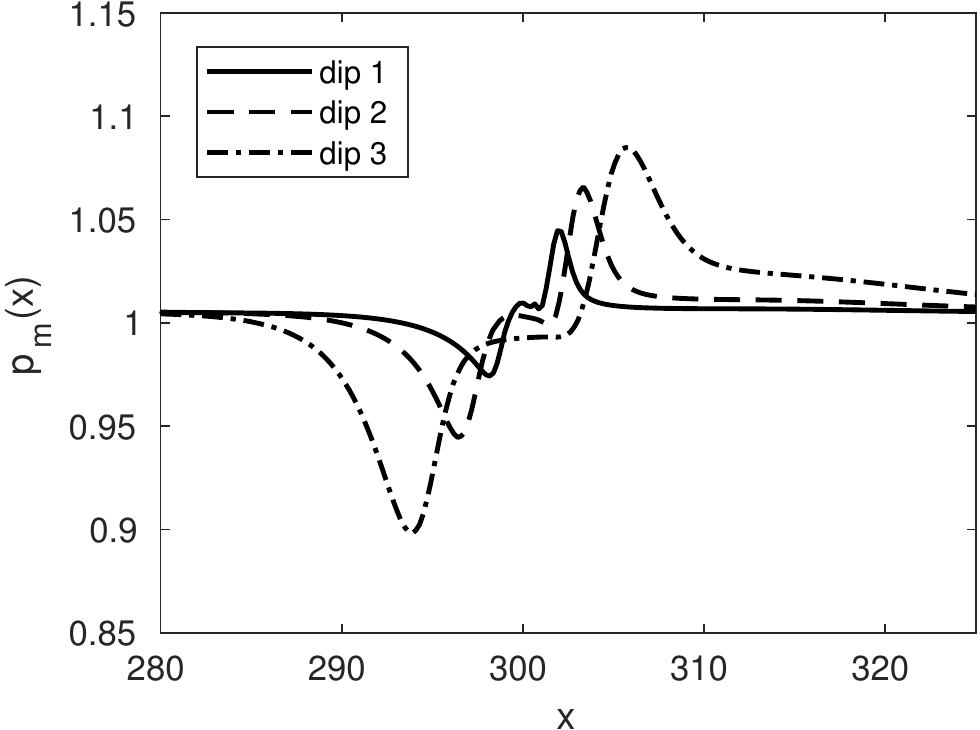}
   \includegraphics[width=6cm]{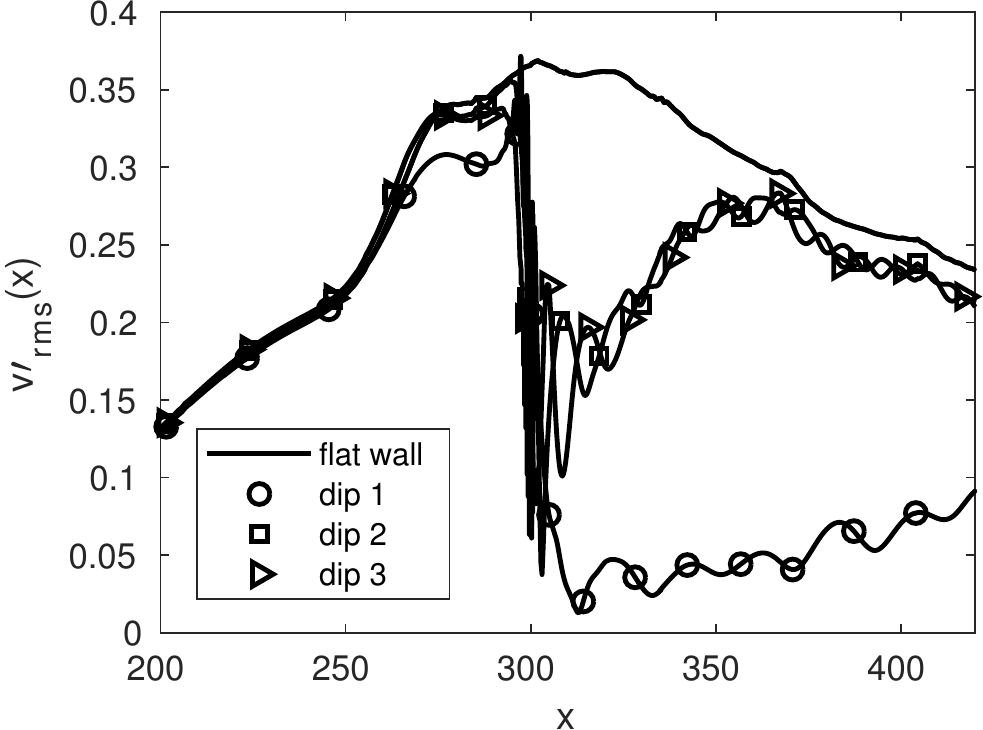} \\
   \hspace{10mm} a) \hspace{78mm} b)
 \end{center}
  \caption{\label{} a) Mean pressure distribution along the wall ($y=1$); b) Root mean square of wall-normal velocity distribution along the wall ($y=0.7$).}
  \label{f23}
\end{figure}

In contrast to the hump deformation, increasing the streamwise length of the dip (see figure \ref{f22}) commensurately affects the mean flow (figures \ref{f23}a). An interesting result is observed in figure \ref{f23}b, where the root mean
square of wall-normal velocity is greatly reduced for the smallest dip, while for the other two extended dips the root mean square is almost the same, but shows an increasing trend for the dip of intermediate streamwise extent (this needs to be further analyzed as a future work).

\begin{figure}[H]
 \begin{center}
    \includegraphics[width=13cm]{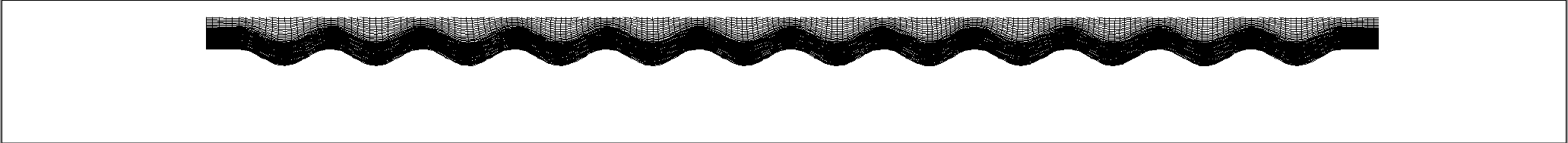} \\
    \includegraphics[width=13cm]{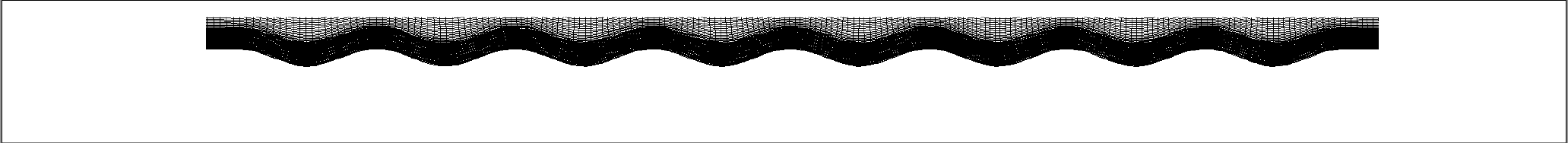} \\
   \includegraphics[width=13cm]{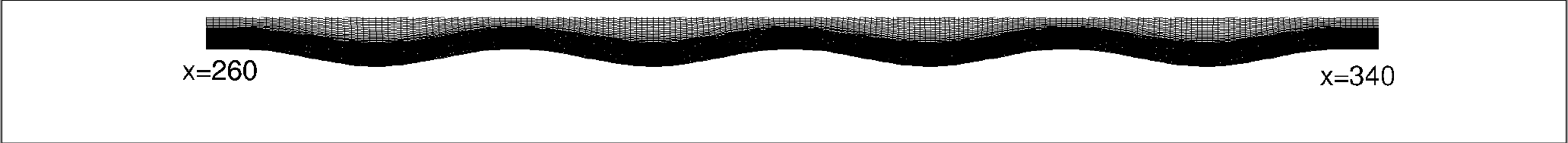}
 \end{center}
  \caption{\label{} Wavy surface type 1 shapes.}
  \label{f24}
\end{figure}

\begin{figure}[H]
 \begin{center}
    \includegraphics[width=6cm]{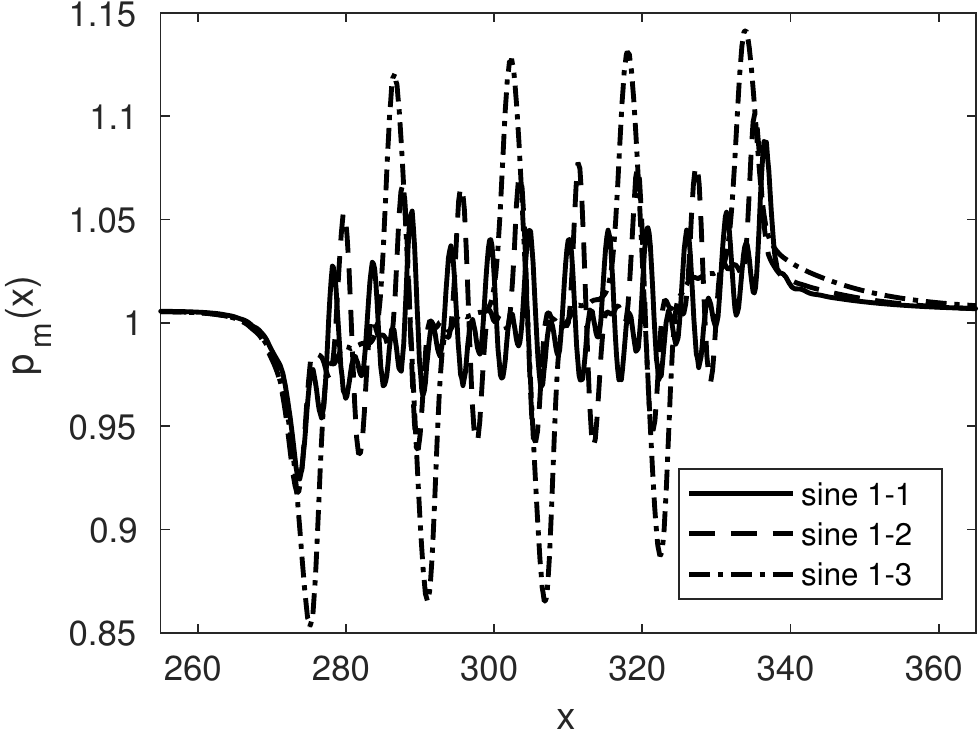}
   \includegraphics[width=6cm]{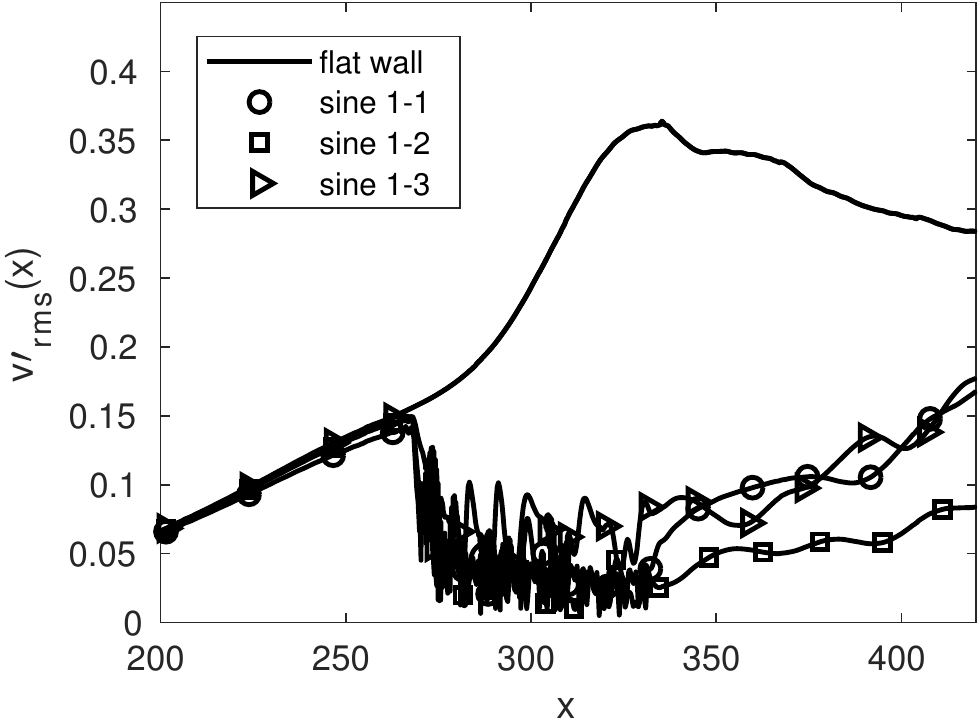} \\
   \hspace{10mm} a) \hspace{78mm} b)
 \end{center}
  \caption{\label{} a) Mean pressure distribution along the wall ($y=1$); b) Root mean square of wall-normal velocity distribution along the wall ($y=0.7$).}
  \label{f25}
\end{figure}

\begin{figure}[H]
 \begin{center}
    \includegraphics[width=6cm]{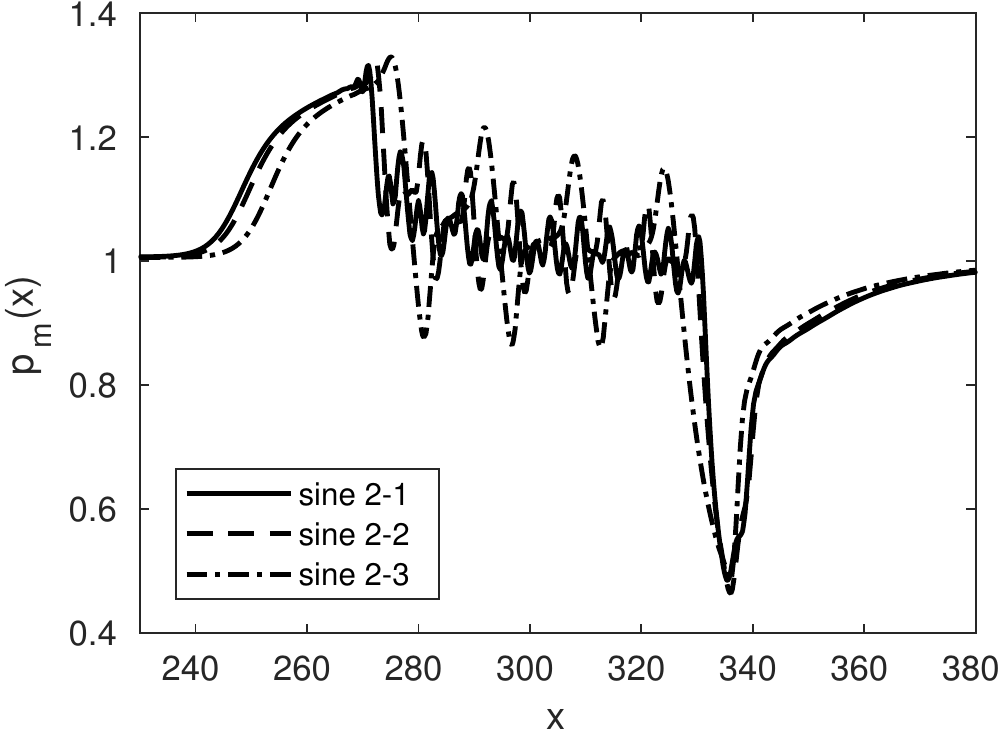}
   \includegraphics[width=6cm]{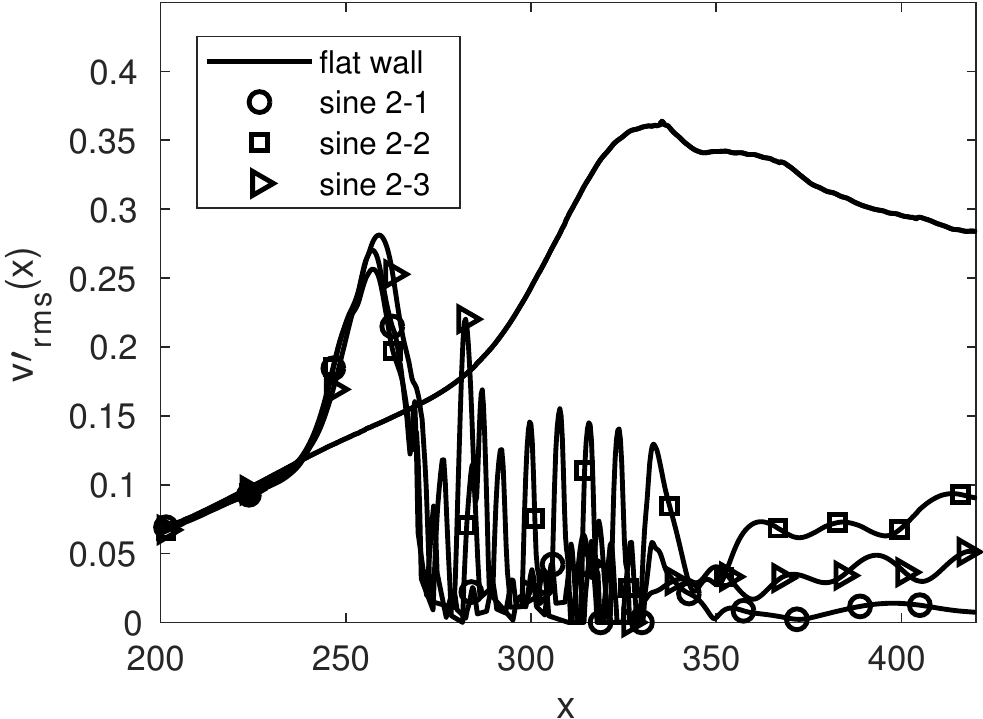} \\
   \hspace{10mm} a) \hspace{78mm} b)
 \end{center}
  \caption{\label{} a) Mean pressure distribution along the wall ($y=1$); b) Root mean square of wall-normal velocity distribution along the wall ($y=0.7$).}
  \label{f26}
\end{figure}

The next two sets of results refer to the wavy surfaces, and seek to determine the effect of varying the wavenumber associated with the sine function used to generate these deformations. The shape of the deformation corresponding to the three wave numbers associated with the wavy surfaces type 1 are shown in figure \ref{f24}, where the middle image is the original wavy surface that was studied in the previous sections (due to similarities, the shapes corresponding to the wavy surface type 2 are not shown). Distributions of the mean pressure in the vicinity of the wall are plotted in figures \ref{f25}a and \ref{f26}a: both show that as the wavenumber is decreased, the mean pressure increases, especially for the wavy surface type 1. There is no significant impact on the disturbance amplitude by changing the wave number of the disturbance, as revealed by distributions of the wall-normal velocity disturbance root mean square in figures \ref{f25}b and \ref{f26}b.

\section{Conclusions}
The response of disturbances propagating inside a high-speed boundary layer to various two-dimensional (2D) surface non-uniformities was studied in this paper, using direct numerical simulations. Two types of disturbances were 
examined: a localized wall pulse in the wall-normal velocity and a periodic wall blowing and suction; both of these disturbances were imposed within the wall boundary condition. The 2D wall non-uniformities included backward and forward steps and their combinations, surface dips, surface humps and two types of wavy surfaces. A study in terms of varying the streamwise width and the height of wall deformations was conducted. Based on the linearized stability analysis, the location of the synchronization point was calculated and found to be located upstream of the wall deformations.
 

Overall, the results showed that the wall deformations have a stabilizing effect on the imposed disturbances, especially for those disturbances that feature an adverse pressure gradient in the upstream followed by a favorable pressure gradient or a succession of adverse and favorable gradients. The effectiveness in reducing the disturbance amplitude of each deformation varied. Various contour plots of pressure disturbance and line plots of 
root mean square of wall-normal velocity disturbance revealed that the forward-backward combination of steps, {\it sine 2} and the forward step were the most effective in 
reducing the amplitudes of both types of disturbances. In contrast, the backward-forward step combination and the dip deformation had the smallest effects on the imposed disturbances, compared to the flat wall. A possible explanation of the mechanism of energy reduction was proposed: part of the energy of the disturbance is deviated outside the boundary layer by the mean flow discontinuity that is generated by the presence of the wall deformation; this is more significant when there is an adverse pressure gradient in the upstream of the deformation.

It was found that the variation of the streamwise width of the wall deformation (while keeping the height constant) plays an important role. It must be mentioned that the variation of the roughness height and the location of the roughness with respect to the synchronization point was studied by Fong et al. \cite{Fong2,Fong3}. For the hump and dip cases, as the length of the deformation was increased, the reduction of the disturbance amplitudes was smaller, especially for the periodic blowing/suction disturbance. This is in contrast to what Fong et al. \cite{Fong3} found, but it must be mentioned that this reduction is valid only for smooth roughness elements, while the non-smooth roughness elements (as the ones considered in Fong et al. \cite{Fong3}), such as the combination between a forward and a backward steps, there is no significant difference by varying the width. For the combination of a backward and forward steps, as the separation length increased the reduction of the energy increased. The variation of the separation length between the forward and  backward steps did not make a difference because the forward step located upstream had the most significant impact on the disturbances.
 

\section{Acknowledgement}

The first author would like to acknowledge partial support from AFORS SFFP Program with Dr. Miguel Visbal as technical monitor, and from NASA EPSCoR RD Program managed by Mississippi State Grant Consortium directed by Dr. Nathan Murray.



\begin{thebibliography}{9}

 \bibitem{Gregory} 
 Gregory, N., Walker, W. S., and Johnson, D. (1951) Part I: The effect on transition of isolated surface excrescenes in the boundary layer; Part II: Brief flight tests on a Vampire I aircraft to determine the effect of isolated surface pimples on transition, {\it Reports and Memoranda 2779}, Aeronautical Research Council.
 
 \bibitem{Drake} 
 Drake, A. and Bender, A. (2009) Surface excrescence transition study, {\it Technical Report AFRL-RB-WP-TR-2009-3109}, Wright-Patterson AFB, OH.
 
 \bibitem{Duncan}  
  Duncan, Crawford, B., Tufts, M.W., Saric, W.S. and Reed, H.L. (2013) Effects of Step Excrescences on Swept-Wing Transition, AIAA Paper 2013-2412.
  
 \bibitem{choudhari1}
Choudhari, M. and Fischer, P. (2005) Roughness induced Transient Growth, AIAA Paper 2005-4765.
  
 \bibitem{Yoon}    
Yoon, S., Barnhardt, M. and Candler, G.V. (2010) Simulations of high-speed flows over an isolated roughness, AIAA Paper 2010-1573.
 
\bibitem{Muppidi}  
Muppidi, S. and Mahesh, K. (2012) Direct numerical simulations of roughness-induced transition in supersonic boundary layers {\it Journal of Fluid Mechanics}, Vol. 693, pp. 288-256.
  
 \bibitem{Iyer}    
 Iyer, P.S., Muppidi, S. and Mahesh, K. (2012) Boundary layer transition in high-speed flows due to roughness, AIAA Paper 2012-1106.
 
 \bibitem{Brehm} 
Brehm, C., Koevary, C., Dackermann, T. and Fasel H.F. (2011) Numerical Investigation of the Influence of Distributed Roughness on Blasius Boundary Layer Stability, AIAA Paper 2011-563.

 \bibitem{Duan}
Duan, L. and Choudhari, M. (2012) Effects of Riblets on Skin Friction in High-Speed Turbulent Boundary Layers, AIAA Paper 2012-1108.

\bibitem{Subbareddy}  
Subbareddy, P.K., Bartkowitcz, M. and Candler, C.V. (2012) Numerical simulations of roughness induced instabilities in the Purdue Mach 6 Tunnel {\it Journal of Fluid Mechanics}, Vol. 748, pp. 848-878.

\bibitem{Rizzetta5} 
Rizzetta, D. P. and Visbal, M. R. (2014) Numerical Simulation of Excrescence Generated Transition, {\it AIAA Journal}, Vol. 52, pp. 385-397.

 \bibitem{Sescu1}
Sescu, A., Visbal and Rizzetta, D. (2014) Numerical Study of Boundary Layer Receptivity to Free-stream Disturbances and Surface Excrescences, {\it AIAA Paper}.

 \bibitem{Sescu5}
Sescu, A., Visbal, M. and Rizzetta, D. (2015) A Study of the Effect of Surface Excrescences and Free-stream Disturbances on Boundary Layers, {\it International Journal for Numerical Methods in Fluids}, Vol. 77, pp. 509-525.

 \bibitem{Chaudhry}
Chaudhry, R.S., Subbareddy, P.K., Nompelis, I. and Candler, G.V. (2015) Direct numerical simulation of roughness-induced transition in the VKI Mach 6 Tunnel, AIAA Paper 2015-0274.

 \bibitem{Fong1}   
Fong, K., Wang, X., and Zhong, X. (2013) Stabilization of Hypersonic Boundary Layer by 2-D Surface Roughness, AIAA Paper 2013-2985.

 \bibitem{Fong2}   
Fong, K., Wang, X., and Zhong, X. (2014) Numerical simulation of roughness effect on the stability of a hypersonic boundary layer, {\it Computers \& Fluids}, Vol. 96, pp. 350-367.

 \bibitem{Fong3}   
Fong, K., Wang, X., and Zhong, X. (2015) Parametric Study on Stabilization of Hypersonic Boundary-Layer Waves Using 2-D Surface Roughness, AIAA Paper 2015-0837.

 \bibitem{Fong4}   
Fong, K., Wang, X., Huang, Y., Zhong, X., McKiernan, G., Fisher, R., and Schneider, S. (2015) Second Mode Suppression in Hypersonic Boundary Layer by Roughness: Design and Experiments, {\it AIAA Journal}, Vol 53 , pp. 3138-3144.

 \bibitem{Duan2}
Duan, L., Wang, X., and Zhong, X. (2013) Stabilization of a Mach 5.92 Boundary Layer by Two-Dimensional Finite-Height
Roughness, {\it AIAA Journal}, Vol. 51, pp. 266-270.

\bibitem{Park} 
Park, D. and Park, S.O. (2016) Study of effect of a smooth bump on hypersonic boundary layer instability, {\it Theoretical and Computational Fluid Dynamics}, May 2016, pp. 1-21.

\bibitem{Bountin} 
Bountin, D., Chimitov, T., Maslov, A., Novikov, A., Egorov, I., Fedorov, A. and Utyuzhnikov, S. (2013) Stabilization of a hypersonic boundary layer using a wavy surface, {\it AIAA Journal}, Vol. 51, pp. 1203-1210.

 \bibitem{Schneider}
 Schneider, P. (2008) Effects of Roughness on Hypersonic Boundary-Layer Transition, {\it Journal of Spacecraft and Rockets}, Vol. 45.

\bibitem{Mack}  
  Mack, L. M. (1975) On the application of linear stability theory and the problem of supersonic boundary-layer transition, {\it AIAA Journal.}, Vol. 13.
  
  \bibitem{Gaponov1}    
Gaponov, S. A. (1993) Excitation of Instability Waves in the Supersonic Boundary Layer by Sound, IUTAM Symposium Potsdam, Springer-Verlag.
  
 \bibitem{Gaponov2}   
Gaponov, S. A. and Smorodsky, B. V. (1999) Supersonic Boundary Layer Receptivity to Streamwise Acoustic Field, IUTAM Symposium, Spinger-Verlag.
   
\bibitem{Fedorov1}    
  Fedorov, A. V., and Khokhlov, A. P. (1991) Excitation of Unstable Modes in a Supersonic Boundary Layer by Acoustic Waves, {\it Fluid Dynamics}, No. 9, pp. 456-467.
  
\bibitem{Fedorov2}     
Fedorov, A. V., and Khokhlov, A. P. (1992) Sensitivity of a Supersonic Boundary Layer to Acoustic Disturbances, {\it Fluid Dynamics}, No. 27, pp. 29-34. 
  
\bibitem{Fedorov3}     
Fedorov, A. V., and Tumin, A. (2011) High-Speed Boundary-Layer Instability: Old Terminology and a New Framework, {\it AIAA Journal.}, Vol. 49, pp. 1647-1657. 
  
\bibitem{Sakaue}   
Sakaue, S., Asia, M., and Nishioka, M. (2000) On the Receptivity process of supersonic laminar boundary layer, Laminar Turbulent Transition, Heidelberg, Springer-Verlag.  
  
 \bibitem{Fedorov3}    
Fedorov, A. V. (2002) Receptivity of High Speed Boundary Layer to Acoustic Disturbances, AIAA Paper 2002-2846.  
  
 \bibitem{Balakumar1}
 Balakumar, P. (2003) Transition in a Supersonic Boundary layer Due to Roughness and Acoustic Disturbances, AIAA Paper 2003-3589.
 
 \bibitem{Balakumar2} 
Balakumar, P. (2005) Transition in a Supersonic Boundary layer Due to Acoustic Disturbances, AIAA Paper 2005-0096.
 
 \bibitem{Balakumar3}
 Balakumar, P. (2009) Receptivity of a supersonic boundary layer to acoustic disturbances, {\it AIAA Journal}, Vol. 47, pp. 1069-178.
 
 \bibitem{Holloway}
Holloway, P. and Sterrett, J., Effect of Controlled Surface Roughness on Boundary-Layer Transition and Heat Transfer
at Mach Numbers of 4.8 and 6.0, TN-D-2054, NASA.

\bibitem{Fujii} 
Fujii, K. (2006) Experiment of the Two-Dimensional Roughness Effect on Hypersonic Boundary-Layer Transition, {\it Journal of Spacecraft and Rockets}, Vol. 43, pp. 731-738.
 
\bibitem{Mortensen}    
Mortensen, C.H. and Zhong, X. (2015) Numerical Simulation of Hypersonic Boundary-Layer Instability in a Real Gas with Two-Dimensional Surface Roughness, AIAA Paper 2015-3077.
 
 \bibitem{liu}
 Liu. X., Osher, S. and Chan, T. (1994) {Weighted essentially non-oscillatory schemes},{\it Journal of Computational Physics}, Vol. 115, pp. 200-212.
 
\bibitem{Tam}  
Tam, C.K.W. and Webb, J.C. (1993), Dispersion-relation-preserving finite difference schemes for Computational Aeroacoustics, {\it Journal of Computational Physics}, Vol. 107, pp. 262-281.

 \bibitem{Bogey3}
Bogey, C., Bailly, C. (2004), A family of low dispersive and low dissipative explicit schemes for flow and noise computation, {\it J. Comp. Phys.}, Vol. 194, pp. 194-214.

\bibitem{Kennedy} 
Kennedy, C.A. and Carpenter, M.H. (1997), Comparison of several numerical methods for simulation of compressible shear layers, {\it NASA Technical Report NASA-97-TP3484}.

\bibitem{Kim}
Kim, J.W. and Lee, D.J. (2000) Generalized Characteristic Boundary Conditions for Computational Aeroacoustics, {\it AIAA Journal}, Vol. 38, pp. 2040-2049.

 \bibitem{Bogey2}
Bogey, C., Bailly, C., and Juve, D. (2002), A shock-capturing methodology based on adaptative spatial filtering
for high-order non-linear computations, {\it Journal of Computational Physics}, Vol. 228, pp. 1447-1465.

\bibitem{Malik}
  Malik, M. R. (1990) Numerical Methods for Hypersonic Boundary Layer Stability, {\it Journal of Computational Physics}, Vol. 86, pp. 376-413.
  
\bibitem{Chuvakhov}
Chuvakhov, P.V. and Fedorov, A.V. (2016) Spontaneous radiation of sound by instability of a highly cooled hypersonic boundary layer, {\it Journal of Fluid Mechanics}, Vol. 805, pp. 188-206.
  
\end{thebibliography}
\end{document}